\documentclass[onecolumn]{revtex4-1}
\usepackage[english]{babel}
\usepackage[latin1]{inputenc}
\usepackage{amsmath}
\usepackage{latexsym}
\usepackage{amssymb}
\usepackage{nameref}
\usepackage{indentfirst}
\usepackage{natbib}
\usepackage{fancyhdr}
\usepackage{graphicx}
\usepackage{epstopdf}
\usepackage{subfig}
\usepackage{lmodern}
\usepackage[T1]{fontenc}
\usepackage{verbatim}
\usepackage{enumerate}
\usepackage{setspace}
\usepackage{color}
\usepackage{bm}
\usepackage{url}
\usepackage{appendix}
\usepackage{paralist}

\newcommand{\be}{\begin{equation}}
\newcommand{\ee}{\end{equation}}

\newcommand{\bea}{\begin{eqnarray}}
\newcommand{\eea}{\end{eqnarray}}

\newcommand{\mO}{\mathcal{O}}
\newcommand{\mA}{\mathcal{A}}
\newcommand{\mD}{\mathcal{D}}
\newcommand{\mF}{\mathcal{F}}
\newcommand{\mM}{\mathcal{M}}

\newcommand{\bs}{\boldsymbol{\sigma}}
\newcommand{\gblue}[1]{{\color{black} #1}}

\begin{document}

\title{Solving the spherical $p$-spin model with the cavity method:\\ equivalence with the replica results}

\author{Giacomo Gradenigo}
\affiliation{Gran Sasso Science Institute, L'Aquila, Italy}
\affiliation{CNR-Nanotec, Institute of Nanotechnology, UOS-Roma, Italy}

\author{Maria Chiara Angelini}
\affiliation{Dipartimento di Fisica, Sapienza Universit\`a di Roma, Italy}

\author{Luca Leuzzi}
\affiliation{CNR-Nanotec, Institute of Nanotechnology, UOS-Roma, Italy}
\affiliation{Dipartimento di Fisica, Sapienza Universit\`a di Roma, Italy}

\author{Federico Ricci-Tersenghi}
\affiliation{Dipartimento di Fisica, Sapienza Universit\`a di Roma, Italy}
\affiliation{CNR-Nanotec, Institute of Nanotechnology, UOS-Roma, Italy}
\affiliation{INFN, Sezione di Roma1, Roma, Italy}

\begin{abstract}

  The spherical $p$-spin is a fundamental model for glassy physics,
  thanks to its analytic solution achievable via the replica method.
  Unfortunately the replica method has some drawbacks: it is very hard
  to apply to diluted models and the assumptions beyond it are not
  immediately clear.  Both drawbacks can be overcome by the use of the
  cavity method, which, however, needs to be applied with care to
  spherical models.

  Here we show how to write the cavity equations for spherical
  $p$-spin models, both in the Replica Symmetric
  (RS) ansatz (corresponding to Belief Propagation) and in the 1-step
  Replica Symmetry Breaking (1RSB) ansatz (corresponding to Survey
  Propagation).  The cavity equations can be solved by a Gaussian (RS)
  and multivariate Gaussian (1RSB) ansatz for the distribution of the
  cavity fields.  We compute the free energy in both ansatzes and
  check that the results are identical to the replica computation,
  predicting a phase transition to a 1RSB phase at low temperatures.

  The advantages of solving the model with the cavity method are many.
  The physical meaning of the ansatz for the cavity marginals is very
  clear. The cavity method works directly with the distribution of
  local quantities, which allows to generalize the method to diluted
  graphs. What we are presenting here is the first step towards the
  solution of the diluted version of the spherical $p$-spin model,
  which is a fundamental model in the theory of random lasers and
  interesting {\it per se} as an easier-to-simulate version of the
  classical fully-connected $p$-spin model.

\end{abstract}

\date{\today}

\maketitle

\tableofcontents

\section{Introduction}
\label{sec:intro}

Spherical models are made of $N$ real variables
$\sigma_i\in\mathbb{R}$ satisfying the global constraint $\sum_i
\sigma_i^2 = N$.  They play a key role among solvable models in
statistical physics, because they usually allow for closed and compact
algebraic solutions~\cite{Berlin52,Baxter82}. Moreover, being the
variables reals, the space of configurations is continuous and
differentiable, thus allowing one to study in these models several kind of
dynamics (e.g. Langevin dynamics or gradient descent like
relaxations). At variance, models whose variables satisfy local
constraints pose more problems. For example, in Ising and Potts models
the variables take discrete values and so the space of configuration
is not continuous; while in $O(n)$ models (e.g.\ with XY or Heisenberg
spins) each variable is continuous, but needs to satisfy a local
constraint of unit norm in an $n$-dimensional space, and this in turn
makes the analytic solution much more complicated, see for
instance~\cite{LRT17,LRT18,LPRT19}.\\

The success of spherical models is well witnessed by the
fully-connected spherical $p$-spin model. For $p\ge3$ this model is
the most used mean-field model for the glassy dynamics. We learned a
lot from it exactly because both the thermodynamics and the dynamics
can be easily
solved~\cite{Crisanti92,Crisanti93,Crisanti95,Bouchaud96}. The
thermodynamic solution has been obtained via the replica method, and
it has a compact analytic form thanks to the spherical constraint: the
solution predicts a random first-order transition from a
high-temperature paramagnetic phase to a low temperature spin glass
phase. The equilibrium and out-of-equilibrium dynamics have been
solved via the generating functional formalism, and it is exact thanks
to the mean-field nature of the model and the spherical
constraint~\cite{Cugliandolo93,Bouchaud96}.

Notwithstanding the success of fully-connected spherical models, we
are well aware they have several unrealistic features:
fully-connectedness is unlikely to happen in any realistic phenomenon
and the spherical constraint is just a global surrogate for the actual
constraint each variable should satisfy locally. In other words, in
realistic models each variable is somehow bounded and one uses the
single global spherical constraint to make computations easier.
Although this approximation is extremely useful, it has some drawbacks. For
example, when the interactions are diluted, a condensation phenomenon
may take place~\cite{Majumdar05,Szavits-Nossan14,GILM19}.

The diluted and sparse versions of a model are particularly
interesting, because moving away from the fully connected limit is
needed in order to study more realistic
phenomena~\cite{Bouchaud04,Biroli06,Biroli08,Biroli2013FragilityMF, Altieri2017Mlayer}.  We reserve the word
\emph{sparse} for graphs with a mean degree $O(1)$, i.e.\ not growing
with $N$, while we use the term \emph{diluted} for a graph which is
not fully-connected, but whose mean degree still grows with $N$.  In
sparse models the couplings do not vanish in the large $N$ limit and
this implies the solution is deeply non-perturbative.  The cavity
method has been developed exactly to solve sparse models \cite{Mezard01}. In very few cases such method has been exploited for fully connected lattices, e.g., in the study of models with discrete variables~\cite{KT95} or in the case of linear interactions, as for the planted SK problem in context of inference~\cite{AKUZ19}.  Diluted models are much less studied in the literature with respect to
fully-connected and sparse models. Nonetheless they are very
interesting for several aspects. They can be used in numerical
simulations as a proxy for fully-connected models which are very
demanding in terms of computing resources.  They appear in models of
random lasers where dilution is induced by the selection rules for the
coupling of light modes in random
media~\cite{Antenucci15a,Antenucci15c,Gradenigo20}. Depending on the
level of dilution, they allow for heterogeneities and local
fluctuations in models that can still be solved similarly to the
fully-connected version, that is, exploiting the fact that couplings are weak
and the graph mean degree diverges.  We believe it is worth dedicating
more efforts in studying the realm of diluted models. In the present
contribution we would like to set up the framework that would allow us
to study diluted models via the cavity method. We are particularly
interested in spherical models, because they are models whose solution
turns out to be particularly simple and compact. However, spherical
models may undergo a condensation transition when the interaction
graph is diluted. How the condensation transition can be avoided in a
$p$-spin model by just modifying the spherical constraint is another
open problem which we are currently investigating and which will be
discussed elsewhere~\cite{AGLR21}. 

The study of whether condensation takes place is a delicate
matter: this depends on a competition between the functional form of
the global constraint, which can even be non-spherical, and the
strength of the interactions, the latter depending on both the order
of the non-linearity and the amount of dilution in the graph. 
Working with Hamiltonian models where variables interact via $p$-body terms and
calling $M=O(N^\delta)$ the number of interaction terms, one would
like to single out the threshold exponent $\delta_c$ such that for
$\delta > \delta_c$ at finite temperature there is no condensation
while for $\delta < \delta_c$ at any temperature the system is in the
condensed phase. So far the situation is clear only for the two
boundaries of the interval of possible values for $\delta$. For
$\delta=p$, which represents the complete graph, condensation is never
found at finite temperature, while the sparse graph, i.e., $\delta=0$,
is always in the condensed phase provided that interactions are
non-linear, i.e., $p>2$. The situation for intermediate values of
$\delta$ is under current investigation, and we expect the present
work to be an important footstep in this direction. For the moment we focus on the
dilution regime where such a condensation phenomenon does not take
place. \\


In the following we present the zero-th order step of the above
program by showing how to use the cavity method to solve the
fully-connected version of spherical spin glass models.  Although the
cavity method is well known \cite{Mezard09}, its use in spherical models did not
appear before in the literature (to the best of our knowledge).  The
application of the cavity method to spherical models is not
straightforward, because one has to decide how to convert a
\emph{global} constraint in a set of \emph{local} ones.  We will
discuss this aspect explicitly and propose a standardized
solution. Once the cavity equations are written, their solution
requires some Ansatz for the distribution of local fields. This is one
of the advantage of the cavity method with respect to the replica
method: all assumptions made in the derivation have a clear and direct
physical meaning. By using a Gaussian Ansatz for the distribution of
local fields (eventually, correlated Gaussian fields in the spin glass
phase where the replica symmetry spontaneously breaks down) we are
able to obtain the exact solution to the spherical $p$-spin spin glass
model, that was previously derived via the replica method. We dedicate the
main text to the derivation of the saddle point equations, to the
illustration of the Ansatz for the local field distributions, to the
discussion on how to implement the spherical constraint and to report
the resulting free-energies. More technical and lengthy derivations,
as the explicit calculations of the free-energy, are postponed to the
Appendices.\\

More in detail: in Sec.~\ref{sec-1} we explain why a Gaussian ansatz for
the cavity marginals is correct in the large degree limit and how to
use it to obtain a closure of the Belief Propagation equations. In particular, in
Sec.~\ref{sec-1:sub-E} we discuss the two possible choices to
implement the spherical constraint in the Belief Propagation
equations, which are equivalent only in the large degree limit.
Sec.~\ref{sec:3} is dedicated to the study of Survey Propagation
equations, i.e., the generalization of Belief Propagation equations in
the case of a one-step-replica-symmetry-breaking scenario. In
Sec.~\ref{Sec:1rsbAnsatz} we present the multivariate Gaussian ansatz
needed for the Survey Propagation equations, recently introduced in Ref. \cite{AKUZ19}, and in Sec.~\ref{sec:3B}
how the explicit closure of the equations is obtained by means of this ansatz. While
the 1RSB expression of the free energy is reported in
Sec.~\ref{sec:1RSB-fe}, its explicit derivation in full detail can be
found in the Appendices.

\section[sec1]{Cavity equations with spherical constraint}
\label{sec-1}

\subsection{Spherical models}

We consider models with $N$ real variables $\sigma_i\in\mathbb{R}$ constrained to satisfy the condition
\be 
\mA[\bm \sigma]\equiv\sum_{i=1}^N\sigma_i^2=N
\label{eq:A0}
\ee
and interacting via $p$-body interactions
\be
\mathcal{H} = - \sum_{a=1}^M J_a \prod_{i\in \partial a} \sigma_i,
\label{eq:H}
\ee
where $\partial a$ is the set of variables entering the $a$-th
interaction and we fix $|\partial a|=p$. If the interaction graph is
fully-connected then $M=\binom{N}{p}$ and the sum runs over all
possible $p$-uples; otherwise, in diluted models, the $M$ interactions
are randomly chosen among the $\binom{N}{p}$ possible $p$-uples.  The
fully-connected versions have been solved via the replica method.

For $p=2$ the model is particularly simple because the energy function
has only two minima and the free-energy can be computed from the
spectrum of the interaction matrix ${\bm J}$.  The model possesses a
spin glass phase at low temperatures, but the replica symmetry never
breaks down and a replica symmetric (RS) ansatz provides the exact
solution \cite{kosterlitz1976spherical}.  In this case the spherical
constraint, although efficient in keeping variables bounded, changes
drastically the low energy physics with respect to models with
e.g. Ising variables: indeed the Sherrington-Kirkpatrick model
\cite{Sherrington75} has a spin glass phase with spontaneous breaking
of the replica symmetry \cite{Parisi80,Parisi83}.

For $p\ge3$ the spherical model is much more interesting since it
undergoes a phase transition to a spin glass phase where the replica
symmetry is broken just once (1RSB phase) \cite{Crisanti92} as in the
analogous model with Ising variables \cite{gardner1985spin}.  More
importantly the thermodynamic phase transition is preceded by a
dynamical phase transition \cite{Crisanti93} which has been connected
to the structural glass transition
\cite{Kirkpatrick87b,Kirkpatrick87c} and to the mode coupling theory
\cite{Goetze09}.  The spherical $p$-spin model with $p\ge3$ represents
now the most used mean-field model for the random first order
transition \cite{castellani2005spin}.

\subsection[sec1-subA]{Self-consistent cavity equations for the local marginals} 
\label{sec-1:sub-A}

The replica method allows to fully characterize the static properties
of the spherical $p$-spin model on complete graphs, as was firstly done in
Ref.~\cite{Crisanti92}.  Our purpose is to study spherical $p$-spin
models, showing that the cavity method is
equivalent to replicas on complete graphs. A complete hypergraph can be seen as a
bipartite graph made of function nodes, representing the interaction
$p$-uplets, and variable nodes, representing the $N$ spins
$\sigma_i$'s. We will indicate the set of links between function and
variable nodes as edges $E$. A complete graph has
$M=\binom{N}{p}=O(N^p)$ function nodes, each of which is linked to $p$
variable nodes. On the other hand, each variable node is linked to
$K=\binom{N-1}{p-1}=O(N^{p-1})$ function nodes.

In order to ensure the extensivity of the energy, not only the $N$
real variables must satisfy the spherical constraint in
Eq.~(\ref{eq:A0}), but the couplings $\{J_a\}$, which are independent
and identically distributed quenched random variables, must be
properly normalized: in the case of symmetric couplings we have
\be
\langle J\rangle = 0 \quad , \quad \langle J^2
\rangle=\frac{p! J_2}{2N^{p-1}}
\label{def:pJ}
\ee
with $J_2=O(1)$ to ensure an extensive energy.  Since we have in mind
to extend the results of the present study to the case of increasing
dilution of the hypergraph, let us start from the statistical ensemble
where the partition function of the model, and hence the corresponding
thermodynamic potentials, is always well defined, i.e., the
\emph{microcanonical ensemble}.

In presence of the spherical constraint written in Eq.~(\ref{eq:A0})
the partition function of the model reads thus
\be \Omega_A(E,N) = \int\prod_{i=1}^N
d\sigma_i~\delta\left(E -\mathcal{H}[\boldsymbol{\sigma}]\right) ~\delta\left(
A-\mA[\boldsymbol{\sigma}]\right).
\label{eq:Z-hard}
\ee
The first, very important, assumption of the present derivation is the
equivalence between the ensemble with hard constraints on both $A$ and
$E$, i.e. the partition function written in Eq.~(\ref{eq:Z-hard}), and
the one where the same spherical constraints are realized via a
Lagrange multiplier. This means that the study of the partition
function in Eq.~(\ref{eq:Z-hard}) is fully equivalent to that of its
Laplace transform:
\be
\mathcal{Z}_\lambda(\beta,N) = \int_{0}^\infty~dA~e^{-\lambda A}~\int_{-\infty}^{+\infty}~dE~e^{-\beta E}~\Omega_A(E,N) =
\int\prod_{i=1}^N d\sigma_i~\exp\left\lbrace -\lambda \sum_{i=1}^N\sigma^2_i + \beta\sum_{a=1}^MJ_a \prod_{i\in \partial a} \sigma_i\right\rbrace,
\label{eq:Z-soft}
\ee
For a given choice of values $A$ and $E$ the ensembles are equivalent
if and only if it is possible to find real values of the Lagrange
multipliers $\lambda$ and $\beta$ such that
\bea
A &=& -\frac{\partial}{\partial\lambda}\log\left[ \mathcal{Z}_\lambda(\beta,N) \right] \nonumber \\
E &=& -\frac{\partial}{\partial\beta}\log\left[ \mathcal{Z}_\lambda(\beta,N) \right]
\label{eq:average-soft}
\eea
In this paper we will consider only choices of $E \propto A \propto N$ 
such that it is possible to find real positive values of $\lambda$ and $\beta$
which solve the equations in Eq.~(\ref{eq:average-soft}).  
It is nevertheless important to keep in mind that there
are situations where Eq.~(\ref{eq:average-soft}) does not have a
solution in terms of either a real $\lambda$ or a real $\beta$: 
this is the situation where the equivalence of ensembles breaks down and we
expect it to happen in sparse hypergraphs, where condensation takes
place. See for instance the recent discussion in~\cite{GILM19}.

Let us now introduce the cavity approach to solve the analyzed
problem.  We will introduce two kinds of cavity messages: with
$\eta_{i\rightarrow a}(\sigma_i)$ we will indicate the
variable-to-function cavity message, that indicates the probability
that the spin on the $i$ node assumes the value $\sigma_i$ in the
absence of the link between the variable node $i$ and the function
node $a$.  Analogously with $\hat{\eta}_{a\rightarrow i}(\sigma_i)$ we
indicate the function-to-variable cavity message.  In the general case
$\eta_{i\rightarrow a}(\sigma_i)$ will depend on all the messages
$\hat{\eta}_{b\rightarrow i}(\sigma_i)$, with $b\in\partial i\setminus
a$, that are correlated random variables. However for tree-like graphs
they are independent, due to the absence of loops.  Loops are
negligible at the leading order also on the Bethe lattice, which is
locally tree-like (there are loops of size $\log(N)$).  A complete
graph is not at all locally tree-like, since each spin participates in
$\mO(N^{p-1})$ interactions, and there are always short
loops. Nevertheless, due to the vanishing intensity of coupling
constants $J_a$, i.e.\ $\langle J^2 \rangle\sim 1/N^{p-1}$,
$\hat{\eta}_{b\rightarrow i}(\sigma_i)$, with $b\in\partial i\setminus
a$, behave as independent random variables even on complete graphs.

This allows us to introduce the following
\emph{cavity equations}:
\bea
\eta_{i\rightarrow a}(\sigma_i) &=&\frac{1}{Z_{i\to a}} \prod_{b \in \partial
  i \setminus a} \hat{\eta}_{b\rightarrow i}(\sigma_i) \label{eq:cavity1}\\ 
\hat{\eta}_{a\rightarrow i}(\sigma_i) & =&\frac{1}{\hat{Z}_{a\to i}} e^{-\frac{\lambda \sigma_i^2}{2 K}} 
\int \prod_{j\in \partial a \setminus i} ~ d\sigma_j ~ \eta_{j\rightarrow a}(\sigma_j) ~ \exp\left\{\beta J_a
  \sigma_i \prod_{j\in \partial a \setminus i}\sigma_j\right\},
\label{eq:cavity2}
\eea
with $Z_{i\to a}$ and $\hat{Z}_{a\to i}$ that are normalization
constants to ensure that the messages are normalized: \be \nonumber
\int_{-\infty}^{\infty}d\sigma_i\eta_{i\to a}(\sigma_i) =
\int_{-\infty}^{\infty}d\sigma_i\hat{\eta}_{i\to a}(\sigma_i)=1 \ .
\ee Let us spend few words on the way we have transformed the global
spherical constraint into the local terms $\exp(-\frac{\lambda
  \sigma_i^2}{2K})$ appearing in the equations for the cavity
marginals $\hat\eta_{a\to i}(\sigma_i)$.  The factor $1/2$ is just
convenient for the definition of the Gaussian distributions (Lagrange
multiplier can be changed by a multiplicative factor without changing
the physics).  Although the most natural place to insert the spherical
constraint would be as an external field in the equation for the
cavity marginal $\eta_{i\to a}(\sigma_i)$, our choice turn out to
simplify the computations and we prove in Sec.~\ref{sec-1:sub-E} to be
equivalent to the other one.  We notice that the idea of moving the
external field from the variables to the interactions is not new. It
is used, for example, in the real space renormalization group.

Once Eqs.~(\ref{eq:cavity1},\ref{eq:cavity2}) are solved (e.g.\ in an
iterative way as in the Belief Propagation algorithm), the local
marginals for each spin ae given by
\be \eta_{i}(\sigma_i) =\frac{1}{Z_{i}}
\prod_{b \in \partial i} \hat{\eta}_{b\rightarrow i}(\sigma_i)
  \label{eq:marginals}
\ee
with $Z_i$ a new normalization constant.

\subsection[sec1-subA]{The Gaussian Ansatz in the large degree limit}

In the fully-connected model, but also in diluted models, the mean
degree grows and diverges in the large $N$ limit. At the same time the
coupling intensities decrease as $N^{-(p-1)/2}$ to ensure well defined
local fields.  In this limit we can close the cavity equations with
the following Gaussian Ansatz for the cavity marginal distribution
\be 
\eta_{i\rightarrow a}(\sigma_i) = \frac{1}{\sqrt{2\pi
    v_{i\rightarrow a}}} \exp\left[-\frac{(\sigma_i-m_{i\rightarrow
      a})^2}{2v_{i\rightarrow a}}\right] 
\propto \exp\left[ \frac{m_{i\to a}}{v_{i \to a}} \sigma_i - \frac{1}{2 v_{i \to a}}\sigma_i^2\right]
\label{eq:Gaussian-ansatz}
\ee
Since $\langle J^2 \rangle \sim 1/N^{p-1}$ the large $N$ limit is
equivalent to as a small $J$ or high-temperature expansion, known as
the Plefka/Georges-Yedidia expansion \cite{Plefka,GY}. Expanding to
second order in $J$, and inserting the Ansatz
Eq. \ref{eq:Gaussian-ansatz}, we get
\begin{align}
\nonumber
\hat\eta_{a \to i}(\sigma_i) &=\frac{1}{\hat{Z}_{a\rightarrow i}} e^{-\frac{\lambda \sigma_i^2}{2 K}} 
\int \prod_{j\in \partial a \setminus i} ~ d\sigma_j ~ \eta_{j\rightarrow a}(\sigma_j) ~ \exp\left\{\beta J_a
  \sigma_i \prod_{j\in \partial a \setminus i}\sigma_j\right\}=\\
\nonumber
&\simeq \frac{1}{\hat{Z}_{a\rightarrow i}}  e^{-\frac{\lambda \sigma_i^2}{2K}}\left[1+\beta J_a \sigma_i \prod_{j \in \partial a \setminus i} m_{j \to a} + \frac{\beta^2 J_a^2}{2} \sigma_i^2 \prod_{j \in \partial a \setminus i}\Big(m_{j\to a}^2 + v_{j \to a}\Big)\right]=\\
&\simeq \frac{1}{\hat{Z}_{a\rightarrow i}}  e^{-\frac{\lambda \sigma_i^2}{2K}}\exp\left\{\beta J_a \sigma_i \prod_{j \in \partial a \setminus
    i} m_{j \to a} + \frac{\beta^2 J_a^2}{2} \sigma_i^2 \left(\prod_{j
      \in \partial a \setminus i}\left(m_{j\to a}^2 + v_{j \to
        a}\right)-\prod_{j \in \partial a \setminus i}m_{j\to
      a}^2\right)\right\}
\label{eq:node-to-spin-A}
\end{align}
and
\begin{align}
\nonumber
\eta_{i \to a}(\sigma_i) &= \frac{1}{Z_{i\rightarrow a}} \prod_{b \in \partial i \setminus a} \hat\eta_{b \to i}(\sigma_i)=\\
&= \frac{1}{Z_{i\rightarrow a}} e^{-\frac{\lambda}{2} \sigma_i^2}\exp\left\{ \beta \sigma_i \sum_{b \in \partial i \setminus a}
  J_b \prod_{j \in \partial b \setminus i} m_{j \to b} +
  \frac{\beta^2}{2} \sigma_i^2 \sum_{b \in \partial i \setminus a}
  J_b^2 \left(\prod_{j \in \partial b \setminus i} \left(m_{j \to b}^2
      + v_{j \to b}\right) - \prod_{j \in \partial b \setminus i} m_{j
      \to b}^2\right) \right\}
\label{eq:spin-to-node-A}
\end{align}
Comparing Eq.~(\ref{eq:Gaussian-ansatz}) and
Eq.~(\ref{eq:spin-to-node-A}), one obtains the following self
consistency equations for the means and the variances of the Gaussian
marginals:
\begin{eqnarray}
\nonumber
\frac{m_{i\to a}}{v_{i \to a}} &=& \beta \sum_{b \in \partial i \setminus a} J_b \prod_{j \in \partial b \setminus i} m_{j \to b}\\
\frac{1}{v_{i \to a}} &=& \lambda - \beta^2 \sum_{b \in \partial i
  \setminus a} J_b^2 \left(\prod_{j \in \partial b \setminus i}
  \left(m_{j \to b}^2 + v_{j \to b}\right) - \prod_{j \in \partial b
    \setminus i} m_{j \to b}^2\right)
\label{eq:closure-cavity}
\end{eqnarray}
The $\lambda$ parameter has to be fixed in order to satisfy the
spherical constraint $\sum_i \langle \sigma_i^2 \rangle = N$, where
the average is taken over the marginals defined in
Eq. (\ref{eq:marginals}).  However, given that we are in a dense
system, cavity marginal and full marginals differ by just terms of
order $O(1/N)$, so we can impose the spherical constraint using cavity
marginals.  These are the replica symmetric cavity equations for dense
(fully-connected or diluted) spherical $p$-spin models.


\gblue{In the limit of large degree (fully-connected or diluted models) the two summations appearing in Eq.~(\ref{eq:closure-cavity}) are over a large number $K$ of terms. So we can use the law of large numbers and the central limit theorem to simplify the self-consistency equations in (\ref{eq:closure-cavity}).
Reminding that in the large $K$ limit the couplings scale according to $\langle J \rangle \sim 1 / K$ and $\langle J^2 \rangle \sim 1 / K$, the second equation in (\ref{eq:closure-cavity}) concentrates $v_{i \to a}$ close its mean value $v=\mathbb{E}(v_{i \to a})$, while the first equation in (\ref{eq:closure-cavity}) implies that the cavity magnetization $m_{i \to a}$ are Gaussian random variables with first moments  $m=\mathbb{E}(m_{i \to a})$ and $q=\mathbb{E}(m_{i \to a}^2)$, satisfying the following equations
\begin{eqnarray}
    \frac{m}{v} &=& \beta \langle J \rangle K \, m^{p-1}  \label{eq:cavity-av-moments-1} \\
    \frac{q}{v^2} &=& \beta^2 \langle J^2 \rangle K \, q^{p-1} + \beta^2 \langle J \rangle^2 K^2 m^{2(p-1)} \label{eq:cavity-av-moments-2} \\
    \frac{1}{v} &=& \lambda - \beta^2 \langle J^2 \rangle K \left((q + v)^{p-1} - q^{p-1}\right) \label{eq:cavity-av-moments-3} 
\end{eqnarray}
By imposing the spherical constraint, $\sum_i \langle \sigma_i^2 \rangle = N$, one gets the identity $q+v=1$ that fixes the Lagrange multiplier and simplifies further the equations
\bea
\lambda &=& \frac{1}{1-q} + \beta^2 \langle J^2 \rangle K
  (1-q^{p-1}) \label{eq:Lagrange-multiplier}\\
m &=& \beta \langle J \rangle K\, m^{p-1}(1-q) \label{eq:closure-cavity-A} \\
q &=& \left[ \beta^2 \langle J^2 \rangle K\, q^{p-1} + \beta^2 \langle J \rangle^2 K^2 m^{2(p-1)}\right](1-q)^2 \label{eq:closure-cavity-B}
\eea
It can be checked by using this expression for $\lambda$ that the normalization of messages $\hat{\eta}_{a\rightarrow i}(\sigma_i)$ is always well defined in the limit of large $N$.


}

\subsection{The replica symmetric free energy}
\label{sec-1:sub-C}

We have now all the pieces we need to compute the replica symmetric
free energy of the model, which is defined as~\cite{Mezard09}:
\be
-\beta F \equiv \beta\left(\sum_{a=1}^{M} \mathbb{F}_a + \sum_{i=1}^N
 \mathbb{F}_i - \sum_{(ai) \in E} \mathbb{F}_{ai}\right)\equiv\sum_{a=1}^{M} \log(Z_a) + \sum_{i=1}^N
\log(Z_i) - \sum_{(ai) \in E}\log(Z_{(ai)}),
\label{eq:free-energy-def}
\ee
where we have respectively
\bea
Z_a &=& \int_{-\infty}^{\infty} \prod_{i\in \partial a} d\sigma_i
~\eta_{i\rightarrow a}(\sigma_i) ~ e^{\beta J_a \prod_{i\in \partial
    a}\sigma_i}  \label{eq:free-energy-Za} \\ 
Z_i &=& \int_{-\infty}^{\infty} d\sigma_i ~ \prod_{a\in \partial i}
\hat{\eta}_{a\rightarrow i}(\sigma_i)  \label{eq:free-energy-Zi} \\ 
Z_{(ai)} &=&  \int_{-\infty}^{\infty} d\sigma_i ~
\hat{\eta}_{a\rightarrow i}(\sigma_i) ~\eta_{i\rightarrow a}(\sigma_i) \label{eq:free-energy-Zai}.
\eea
The computation of these three terms is reported in the Appendix
\ref{app:RSFreeEnergy}. Here we just report the final result:
\bea
-\beta F_{\textrm{RS}} &=&
\frac{N}{2}\left[\frac{\beta^2}{2} (1-q^p) J_2 + \log(1-q)+ \frac{q}{(1-q)}\right],
\label{eq:frs-final}
\eea
The free energy written in Eq.~(\ref{eq:frs-final}) is identical to
that of the spherical $p$-spin computed with replicas in the replica
symmetric case, see Eq.~(4.4) of~\cite{Crisanti92}. From now on we
will set $J_2=1$.\\

\subsection[sec1-subE]{Alternatives for the spherical constraint: equivalence in the large-$N$ limit.} 
\label{sec-1:sub-E}

The experienced reader will have probably noticed that the way we have
introduced the spherical constraint in the cavity equations is not,
perhaps, the most natural one, that would correspond to an
\emph{external field} of intensity $\lambda$ acting on every spin. As
such, we should have put
\be
\eta_{i\rightarrow a}(\sigma_i) ~\propto~e^{-\frac{\lambda}{2}\sigma_i^2},
\label{eq:spherical-eta}
\ee
rather than
\be
\hat{\eta}_{a\rightarrow i}(\sigma_i) ~\propto~e^{-\frac{\lambda}{2K}\sigma_i^2},
\label{eq:spherical-eta-hat}
\ee
as we have done in the equations for the cavity marginals,
Eq.~(\ref{eq:cavity1}) and Eq.~(\ref{eq:cavity2}). In what follows we
show that the choice of where to put the spherical constraint is
arbitrary in the large-$N$ limit. In practice we are going to show
that either we let the constraint act as an external field in the
\emph{variable-to-function} message $\eta_{i\rightarrow a}(\sigma_i)$,
as in Eq.~(\ref{eq:spherical-eta}), or inside the
\emph{function-to-variable} marginal $\hat\eta_{a\rightarrow
  i}(\sigma_i)$, as in Eq.~(\ref{eq:spherical-eta-hat}), in both cases
we obtain the same expression for the free energy to the leading order
in $N$. The reader must therefore bare in mind that the two ways to
put the constraint in the cavity equations \emph{might not be
  equivalent} in the case of a graph with finite connectivity.

After a trial and error procedure we realized that the choice in
Eq.~(\ref{eq:spherical-eta-hat}) makes all calculations simpler, so
that we opted for this one. We have already shown that by doing so we
obtain, at high temperature, a free energy which is identical to the
one obtained from mean-field replica
calculations,~Eq.~(\ref{eq:frs-final}).  We now want to show
explicitly that, term by term and beside any further assumption as the
one of homogeneity, the free energy in the high temperature ergodic
phase is identical for the two choices [Eq.~(\ref{eq:spherical-eta})
  and Eq.~(\ref{eq:spherical-eta-hat})] to introduce the constraint.

Let us term $\eta_{i\rightarrow a}^{(\lambda)}(\sigma_i)$ and
$\hat{\eta}_{a\rightarrow i}^{(\lambda)}(\sigma_i)$ the local cavity
marginals corresponding to the case where the \emph{field} $\lambda$
acts directly on the spin:
\bea
\eta_{i\rightarrow a}^{(\lambda)}(\sigma_i) &=&\frac{1}{Z_{i\to a}^{(\lambda)}} e^{-\frac{\lambda \sigma_i^2}{2}} \prod_{b \in \partial
  i \setminus a} \hat{\eta}_{b\rightarrow i}^{(\lambda)}(\sigma_i) \label{eq:cavity1-lambda}\\ 
\hat{\eta}_{a\rightarrow i}^{(\lambda)}(\sigma_i) & =&\frac{1}{\hat{Z}_{a\to i}^{(\lambda)}} 
\int_{-\infty}^{\infty} \prod_{j\in \partial a \setminus i} ~ d\sigma_j ~ \eta_{j\rightarrow a}^{(\lambda)}(\sigma_j) ~ \exp\left\{\beta J_a
  \sigma_i \prod_{j\in \partial a \setminus i}\sigma_j\right\}.
\label{eq:cavity2-lambda}
\eea
Accordingly, since in the function-to-variable messages there is now
no trace of the external field, one has to consider the following
modified definition of the entropic term in the local partition
functions:
\bea
Z_a^{(\lambda)} &=& \int_{-\infty}^{\infty} \prod_{i\in \partial a} d\sigma_i
~\eta_{i\rightarrow a}^{(\lambda)}(\sigma_i) ~ e^{\beta J_a \prod_{i\in \partial
    a}\sigma_i}  \label{eq:free-energy-Za-lambda} \\ 
Z_i^{(\lambda)} &=& \int_{-\infty}^{\infty} d\sigma_i ~e^{-\lambda \sigma_i^2 /2}~ \prod_{a\in \partial i}
\hat{\eta}_{a\rightarrow i}^{(\lambda)}(\sigma_i) \label{eq:free-energy-Zi-lambda} \\ 
Z_{(ai)}^{(\lambda)} &=&  \int_{-\infty}^{\infty} d\sigma_i ~
\hat{\eta}_{a\rightarrow i}^{(\lambda)}(\sigma_i) ~\eta_{i\rightarrow a}^{(\lambda)}(\sigma_i) \label{eq:free-energy-Zai-lambda}.
\eea
Our task is now to show that:
\be
\sum_{a=1}^{M} \log(Z_a) + \sum_{i=1}^N
\log(Z_i) - \sum_{(ai) \in E}\log(Z_{(ai)}) = \sum_{a=1}^{M} \log(Z_a^{(\lambda)}) + \sum_{i=1}^N
\log(Z_i^{(\lambda)}) - \sum_{(ai) \in E}\log(Z_{(ai)}^{(\lambda)}).
\label{eq:identity-free-energies}
\ee
The key observation is that, in order to have overall consistency, the
Gaussian ansatz for the variable-to-function message \emph{must} be
the same in both cases, that is:
\be
\eta_{i\rightarrow a}(\sigma_i) =  \frac{1}{\sqrt{2\pi
    v_{i\rightarrow a}}} \exp\left[-\frac{(\sigma_i-m_{i\rightarrow
      a})^2}{2v_{i\rightarrow a}}\right] = \eta_{i\rightarrow a}^{(\lambda)}(\sigma_i).
\label{eq:equivalence-eta}
\ee
The assumption of Eq.~(\ref{eq:equivalence-eta}) allows us to conclude
immediately that $Z_a^{(\lambda)} = Z_a$, so that the identity we need
to prove reduces to:
\be
\sum_{i=1}^N \log(Z_i) - \sum_{(ai) \in E}\log(Z_{(ai)}) =  \sum_{i=1}^N
\log(Z_i^{(\lambda)}) - \sum_{(ai) \in E}\log(Z_{(ai)}^{(\lambda)})
\label{eq:identity-free-energies-2nd}
\ee
By exploiting Eq. (\ref{eq:equivalence-eta}) once again we obtain
\be
\hat{\eta}_{a\rightarrow i}^{(\lambda)}(\sigma_i) =\frac{1}{\hat{Z}_{a\to i}^{(\lambda)}} 
\int_{-\infty}^{\infty} \prod_{j\in \partial a \setminus i} ~ d\sigma_j ~ \eta_{j\rightarrow a}(\sigma_j) ~ \exp\left\{\beta J_a
  \sigma_i \prod_{j\in \partial a \setminus i}\sigma_j\right\},
\ee
that, by comparison with the defintion Eq.~(\ref{eq:cavity2}) leads to
\be
\hat{\eta}_{a\rightarrow i}^{(\lambda)}(\sigma_i)~\hat{Z}_{a\to i}^{(\lambda)} = \hat{\eta}_{a\rightarrow i}(\sigma_i)~\hat{Z}_{a\to i}~e^{\frac{\lambda \sigma_i^2}{2K}}
\label{eq:nontrivial-identity},
\ee
so that
\be
\hat{\eta}_{a\rightarrow i}^{(\lambda)}(\sigma_i) = \hat{\eta}_{a\rightarrow i}(\sigma_i)~\frac{\hat{Z}_{a\to i}}{\hat{Z}_{a\to i}^{(\lambda)}}~e^{\frac{\lambda \sigma_i^2}{2K}}
\label{eq:nontrivial-identity2}.
\ee
By inserting Eq.~(\ref{eq:nontrivial-identity2}) in the definition of
$Z_i^{(\lambda)}$ in Eq.~(\ref{eq:free-energy-Zi-lambda}) one finds:
\bea Z_i^{(\lambda)} &=& \int_{-\infty}^{\infty} d\sigma_i~e^{-\lambda
  \sigma_i^2 /2}~\prod_{a\in \partial i} \hat{\eta}_{a\rightarrow
  i}^{(\lambda)}(\sigma_i) \nonumber \\
&=&\prod_{a\in \partial i} \left(\frac{\hat{Z}_{a\to i}}{\hat{Z}_{a\to i}^{(\lambda)}}\right) \int_{-\infty}^{\infty} d\sigma_i~\prod_{a\in \partial i} \hat{\eta}_{a\rightarrow i}(\sigma_i)
\nonumber \\
&=& \prod_{a\in \partial i}\left(\frac{\hat{Z}_{a\to i}}{\hat{Z}_{a\to i}^{(\lambda)}}\right) ~Z_i, 
\eea
so that the identity that we want prove is further simplified in
\be
\sum_{(ai) \in E}\log(Z_{(ai)}) =  \sum_{(ai) \in E}\log(Z_{(ai)}^{(\lambda)}) - \sum_{i=1}^N \sum_{a\in\partial i } \log\left( \frac{\hat{Z}_{a\to i}}{\hat{Z}_{a\to i}^{(\lambda)}} \right).
\label{eq:identity-free-energies-3rd}
\ee
Using, once again, 
Eq.~(\ref{eq:equivalence-eta}) we can  write
\bea
Z_{(ai)}^{(\lambda)} &=& \int_{-\infty}^{\infty} d\sigma_i ~
\hat{\eta}_{a\rightarrow i}^{(\lambda)}(\sigma_i) ~\eta_{i\rightarrow a}^{(\lambda)}(\sigma_i) \nonumber \\
&=& \int_{-\infty}^{\infty} d\sigma_i ~
\hat{\eta}_{a\rightarrow i}^{(\lambda)}(\sigma_i) ~\eta_{i\rightarrow a}(\sigma_i) \nonumber \\
& = & \frac{\hat{Z}_{a\to i}}{\hat{Z}_{a\to i}^{(\lambda)}} ~\int_{-\infty}^{\infty} d\sigma_i ~
e^{\frac{\lambda \sigma_i^2}{2K}}~\hat{\eta}_{a\rightarrow i}(\sigma_i) ~\eta_{i\rightarrow a}(\sigma_i) \nonumber \\
&\simeq& \frac{\hat{Z}_{a\to i}}{\hat{Z}_{a\to i}^{(\lambda)}} ~Z_{(ai)},
\label{eq:normalization-identity}
\eea
where the last line equality holds for large $N$ (see
Eqns.~(\ref{eq:Zai-compute-first}), (\ref{eq:Zai-compute-2nd}) and
(\ref{eq:Zai-compute-3rd}) in Appendix A). The $N\rightarrow\infty$ limit is
equivalent to the $K\rightarrow\infty$ limit, since $K \sim
N^{p-1}$. Finally, by plugging the result of
Eq.~(\ref{eq:normalization-identity}) into
Eq.~(\ref{eq:identity-free-energies-3rd}) we can conclude that the
identity in Eq.~(\ref{eq:identity-free-energies-3rd}) is true in the
limit $N\rightarrow\infty$. We have thus demonstrated that in the
large-$N$ limit it is equivalent, and thus just a matter of
convenience, to write down explicitly the spherical constraint inside
the definition of the function-to-variable message
$\hat{\eta}_{a\rightarrow i}(\sigma)$, as we have done, or inside the
definition of the variable-to-function one, $\hat{\eta}_{i\rightarrow
  a}(\sigma)$.

\section{One step Replica Symmetry Breaking solution}
\label{sec:3}

In the previous sections we have reviewed the replica symmetric
solution that is the stable one for high temperatures. In this phase
we have written closed cavity equations for the marginal distributions
of the variables, relying on the assumption that the joint
distribution of the cavity variables is factorized as in a single pure state.

However, lowering the temperature, it is known from the replica solution \cite{Crisanti92}, that several metastable glassy states arise on top of the paramagnetic state. Their number being exponential in $N$ with a rate $\Sigma$ called \emph{complexity}. The function $\Sigma(f)$ is in general an increasing function of the state free-energy $f$, with a downwards curvature (for stability reasons as for the entropy).

Comparing the total free-energy of the glassy states computed using $\Sigma(f)$ and the paramagnetic free-energy \cite{Zamponi} one can derive the dynamical critical temperature $T_d$ where the ergodicity breaks down and the thermodynamic critical temperature, also called Kauzmann temperature $T_K$, where a phase transition to a replica symmetry breaking phase takes place.

Below $T_d$ the dynamics of the model is dominated by the states of larger free-energy, so-called threshold states, which are the most abundant and always exponentially many in $N$ (although a more refined picture has been recently presented in \cite{folena2019memories}).

For $T<T_d$ the Gibbs measure is split over many different states, such that two different equilibrium configurations
can be in the same (metastable) state or in different states.  Defining
the overlap between two different configurations as how much they are
close to each other, the 1RSB phase is characterized by an overlap
$q_1$ between configurations inside the same pure state (independently
of the pure state) and an overlap $q_0<q_1$ between configurations in
two different states.

In formulas, the presence of many metastable pure states yields an
additional contribution to the free-energy. The complexity
$\Sigma(f)$, that counts the number of ``states'' (disjoint ergodic
components of the phase-space) with the same free-energy $f$ can be written as
\be
\Sigma(f) = \frac{1}{N}\log\left[\sum_{\eta=1}^{\mathcal N}~\delta(f-f_\eta)\right] \ , 
\label{eq:Sc}
\ee
where $\mathcal N$ is the total number of metastable glassy states
(formally they can be defined as the non-paramagnetic stationary
points of the TAP free-energy \cite{Thouless77}) and $f_\eta$ is the
free energy of the glassy state $\eta$.  Please notice that expression
in Eq.~(\ref{eq:Sc}) is identical to the standard microcanonical
definition of entropy, with the only difference that now we measure
the number of phase-space regions with the same free-energy rather
than the volume of phase space with the same energy.  The total
free-energy is thus given by:
\be \mF=-\frac{1}{\beta N}\log
Z=-\frac{1}{\beta N}\log\left( \sum_{\eta}e^{-\beta N
  f_{\eta}}\right)= -\frac{1}{\beta N}\log\int df
\sum_{\eta}~\delta(f-f_\eta) e^{-\beta N f}=-\frac{1}{\beta N}\log\int
df e^{-N(\beta f-\Sigma)} \ee

The problem is that we do not know how to characterize the different
states and how to count them to obtain $\Sigma$: we are still not able
to compute $\mF$.  In the following we will solve this problem
applying the method of real coupled replicas introduced by Monasson in
Ref. \cite{Monasson95} (see also \cite{Mezard09} for a rigorous
derivation and \cite{Zamponi} for a pedagogical review).  This method
was applied to the spherical $p$-spin in Ref.~\cite{Mezard99b} to
compute the 1RSB free-energy with a replica computation.  The idea of
Ref.~\cite{Monasson95} is to introduce $x$ real clones, that we will
call replicas, on a single realization of a graph. These replicas will
be infinitesimally coupled together in such a way that, even when the
coupling between them goes to zero, they will all fall in the same
pure state below $T_d$: this cloning method is a way to select a state
equivalent to what is usually done in ferromagnetic systems to select
a state adding an infinitesimal magnetic field.  The free energy
$\Phi(x)$ of $x$ replicas in the same state is:
\be
\Phi(x)=-\frac{1}{\beta N}\log\left( \sum_{\eta}e^{-\beta N
  xf_{\eta}}\right)=-\frac{1}{\beta N}\log\int df e^{-N(\beta x
  f-\Sigma)}=-\frac{1}{\beta}\max_f (\beta x f-\Sigma(f)).
\ee

The complexity in this way simply results in the Legendre transform of
the free energy of the replicated system.  The total free-energy in
the 1RSB phase is derived passing to the analytical continuation of
$x$ to real values and turns out to be: $\mF=\min_x
\frac{\Phi(x)}{x}$. Beside the Monasson-Mezard clonig method, which is mostly useful to study the complexity of systems without quenched disorder, it is worth recalling the physical meaning of the analytic continuation to positive real values of $x$ in a more general setting: it allows to compute the large deviations of the free-enenergy, e.g., its sample-to-sample fluctuations~\cite{CPSV92,PDGR19}.

In the following we will use this cloning method to write 1RSB closed
cavity equations for the spherical $p$-spin, in a way analogous to
what has been done in Ref.~\cite{AKUZ19} for the planted SK model.  In
a situation with many pure states, the factorization of the
distribution of the cavity variables is valid only inside a single
pure state: we can thus still write some cavity closed equations
considering the coupled replicas in a same pure state.  Then, we will
compute the 1RSB free energy in a cavity approach below $T_d$,
obtaining exactly the same expression found with replica computations
in Refs.~\cite{Crisanti92,Mezard99b}.

\subsection{The ansatz for the distribution of $x$ coupled replicas}
\label{Sec:1rsbAnsatz}

For the RS phase, in the dense case, we have written a Gaussian ansatz
for the marginal probability of the spin on a given site in
Eq.~(\ref{eq:Gaussian-ansatz}).  In the 1RSB phase, we will consider
the \emph{joint probability distribution} of $x$ coupled replicas that
are all in a same pure state. We will comment in the next sections on
the choice and the physical meaning of $x$.  In order to lighten the
notation, let us indicate as $\boldsymbol{\sigma}_i=\lbrace
\sigma_i^\alpha \rbrace$, $\alpha =1,\ldots,x$, the vector of all $x$
replicas on site $i$.  The 1RSB form of the ansatz for the marginal
probability $\eta_{i \rightarrow a}(\boldsymbol{\sigma_i})$ amounts to
\be
\eta_{i\rightarrow a}(\boldsymbol{\sigma}_i) = \int_{-\infty}^\infty dm_{i\rightarrow a}~\frac{1}{\sqrt{2\pi\Delta_{i\rightarrow a}^{(0)}}}\exp\left( -\frac{(m_{i\rightarrow a}-h_{i\rightarrow a})^2}{2\Delta_{i\rightarrow a}^{(0)}}\right)
\frac{1}{\left[ \sqrt{2\pi\Delta_{i\rightarrow a}^{(1)}}\right]^x} \exp\left(-\sum_{\alpha=1}^x\frac{(\sigma_i^\alpha-m_{i\rightarrow a})^2}{2\Delta_{i\rightarrow a}^{(1)}} \right).
\label{eq:1RSB-ansatz}
\ee
This 1RSB Ansatz was firstly introduced in Ref.~\cite{AKUZ19}.
By shortening the integration measure for the joint probability
distribution $\boldsymbol{\sigma}_i$ with the symbol
\be
\int \mD \bs_i = \int_{-\infty}^\infty\prod_{\alpha =1}^x d\sigma_i^\alpha, 
\ee
and defining the distribution
\bea
Q_{i\to a}\left(m_{i\rightarrow a} \right) \equiv
\frac{1}{\sqrt{2\pi\Delta_{i\rightarrow a}^{(0)}}} \exp\left( -\frac{(m_{i\rightarrow a}-h_{i\rightarrow a})^2}{2\Delta_{i\rightarrow a}^{(0)}} \right),
\label{eq:Q-1rsb-ansatz}
\eea
the first diagonal and second moments of the cavity marginal are simply computed as:
\bea
\langle \sigma_i^\alpha \rangle &=& \int \mD \bs_i ~\sigma_i^\alpha~\eta_{i\rightarrow a}(\boldsymbol{\sigma}_i)=
\int_{-\infty}^\infty dm_{i\rightarrow a}~m_{i\rightarrow a}~
Q_{i\to a}\left(m_{i\rightarrow a} \right)
= h_{i\rightarrow a} \nonumber \\
\langle (\sigma_i^\alpha)^2 \rangle &=& \int \mD \bs_i ~(\sigma_i^\alpha)^2~\eta_{i\rightarrow a}(\boldsymbol{\sigma}_i) =
\int_{-\infty}^\infty dm_{i\rightarrow a}~(\Delta_{i\rightarrow a}^{(1)}+m_{i\rightarrow a}^2)~Q_{i\to a}\left(m_{i\rightarrow a} \right)
=
\Delta_{i\rightarrow a}^{(1)} + \Delta_{i\rightarrow a}^{(0)} + h_{i\rightarrow a}^2  \nonumber \\
\langle \sigma_i^\alpha \sigma_i^\beta \rangle &=& \int \mD \bs_i ~\sigma_i^\alpha~\sigma_i^\beta~\eta_{i\rightarrow a}(\boldsymbol{\sigma}_i)=
\int_{-\infty}^\infty dm_{i\rightarrow a}~m_{i\rightarrow a}^2~
Q_{i\to a}\left(m_{i\rightarrow a} \right)
=
\Delta_{i\rightarrow a}^{(0)} + h_{i\rightarrow a}^2
\label{eq:1RSB-distribution-moments}
\eea

Let us comment briefly on the form of the Ansatz. The marginal
probability of a single replica in a given state is still a Gaussian,
being on a dense graph.  If the real replicas are coupled, they will
fall in the same state. The only effect of the infinitesimal coupling
between the replicas will be that the configurations of the real
replicas will be independent variables extracted from the same
distribution in each state, once average $m$ and variance
$\Delta^{(1)}$ are given:
\be
\eta^{\rm s}_{i\rightarrow a}\left(\sigma_i^\alpha \right) \equiv \frac{1}{\sqrt{2\pi\Delta_{i\rightarrow a}^{(1)}}}
\exp\left( -\frac{(\sigma_i^\alpha-m_{i\rightarrow a})^2}{2\Delta_{i\rightarrow a}^{(1)}}\right).
\label{eq:replicas-1rsb-ansatz}
\ee
In the same way, the average magnetizations in different states will
be independent variables extracted from the same distribution $Q_{i\to
  a}(m_{i\rightarrow a}) $, that will depend on $\Delta^{(0)}$ and
$h$, see, e.g., ref. \cite{Mezard85}.

With this simple scenario in mind, we can give a simple physical
interpretation to the parameters of the distribution in
Eq.~(\ref{eq:1RSB-ansatz}) rewriting them as:
\bea
h_{i\rightarrow a} &=& \langle \sigma_i^\alpha \rangle \nonumber \\
\Delta^{(1)}_{i\rightarrow a} &=& \langle (\sigma_i^\alpha)^2 \rangle  - \langle \sigma_i^\alpha \sigma_i^\beta \rangle = 1- q_1^{i\rightarrow a}\nonumber \\
\Delta^{(0)}_{i\rightarrow a} &=& \langle \sigma_i^\alpha \sigma_i^\beta \rangle - \langle \sigma_i^\alpha \rangle^2 = q_1^{i\rightarrow a} - q_0^{i\rightarrow a},
\label{eq:1RSB-parameters}
\eea
where the average is taken with respect to the probability
distribution in Eq.~(\ref{eq:1RSB-ansatz}).  $q_1^{i\rightarrow a}$
and $q_0^{i\rightarrow a}$ are the local overlap (in absence of the
link from $i$ to $a$) inside a state and between states that we
mentioned at the beginning of this Section.  Obviously on a complete
graph, they will be independent of $i$ and $a$, as for the only
parameter in the RS case (the magnetization) in the homogeneous
case. However, we here prefer to write explicitly the dependence on
$i$ and $a$, because in this way the equations we will obtain could be
easily applied to non-complete graphs.

\subsection{1RSB cavity equations}
\label{sec:3B}

We now write the replicated cavity equations for the 1RSB Ansatz
introduced in the previous section:
\bea \eta_{i\rightarrow a}(\boldsymbol{\sigma}_i) &\propto& \prod_{b
  \in \partial i \setminus a} \hat{\eta}_{b\rightarrow
  i}(\boldsymbol{\sigma}_i) \label{eq:cavity1-1RSB}
\\ \hat{\eta}_{a\rightarrow i}(\boldsymbol{\sigma}_i) &\propto&
e^{-\frac{\lambda}{2 K} \sum_{\alpha=1}^x (\sigma^{\alpha}_i)^2}
\int_{-\infty}^{\infty} \prod_{k\in \partial a \setminus i} ~
\mathcal{D}\boldsymbol{\sigma}_k ~ \eta_{k\rightarrow a}(\boldsymbol{\sigma}_k) ~
\exp\left\{\beta J_a \sum_{\alpha=1}^x \sigma_i^\alpha \prod_{k\in \partial a \setminus
  i}\sigma_k^\alpha \right\}.
\label{eq:cavity2-1RSB}
\eea

We have omitted the normalization factors that are irrelevant in the
subsequent computations.  As we did for the RS case, in the dense
limit we take the leading term in a small $J_a$ expansion (valid in
the large $N$ limit for dense graphs) and in this setting we will
close the equations on the parameters of the multivariate
Gaussian. That is, we write:
\bea
& & \int_{-\infty}^{\infty} \prod_{k\in \partial a \setminus i} ~
\mathcal{D}\boldsymbol{\sigma}_k ~ \eta_{k\rightarrow a}(\boldsymbol{\sigma}_k) ~
~\exp\left\{\beta J_a \sum_{\alpha=1}^x \sigma_i^\alpha \prod_{k\in \partial a \setminus
  i}\sigma_k^\alpha \right\} \simeq \nonumber \\
~&\simeq & \int_{-\infty}^{\infty} \prod_{k\in \partial a \setminus i} ~
\mathcal{D}\boldsymbol{\sigma}_k ~ \eta_{k\rightarrow a}(\boldsymbol{\sigma}_k) ~ \left[ 1 + \beta J_a \sum_{\alpha=1}^x \sigma_i^\alpha \prod_{k\in \partial a \setminus i} \sigma_k^\alpha + \frac{1}{2} \beta^2 J^2_a~\left( \sum_{\alpha=1}^x \sigma_i^\alpha \prod_{k\in \partial a \setminus i} \sigma_k^\alpha\right)^2 \right] = \nonumber \\
~&=& 1 + \beta J_a \left(\prod_{k\in \partial a \setminus i} h_{k\rightarrow a}\right)~~\sum_{\alpha=1}^x \sigma_i^\alpha  +
\frac{1}{2} \beta^2 J^2_a~\left[\prod_{k\in \partial a \setminus i} \left( \Delta_{k\rightarrow a}^{(1)} + \Delta_{k\rightarrow a}^{(0)} + h_{k\rightarrow a}^2 \right)\right]~~\sum_{\alpha=1}^x \left(\sigma_i^\alpha\right)^2 + \nonumber \\
 && + \frac{1}{2} \beta^2 J^2_a~\left[\prod_{k\in \partial a \setminus i} \left(\Delta_{k\rightarrow a}^{(0)} + h_{k\rightarrow a}^2\right)\right]~~\sum_{\alpha\neq\beta}^x \sigma_i^\alpha\sigma_i^\beta \simeq \nonumber \\
~&\simeq&~\exp\left\lbrace \hat{A}_{a\to i}~\sum_{\alpha=1}^x \sigma_i^\alpha -\frac{1}{2}~\hat{B}_{a\to i}^{(\text{d})}~\sum_{\alpha=1}^x (\sigma_i^\alpha)^2 + \frac{1}{2}~\hat{B}_{a\to i}^{(\text{nd})}~ \sum_{\alpha\neq\beta}^x \sigma_i^\alpha\sigma_i^\beta \right\rbrace,
\eea
where the three coefficients are respectively
\bea
\hat{A}_{a\rightarrow i} &=& \beta J_a \prod_{k\in \partial a \setminus i} h_{k\rightarrow a} \nonumber \\
\hat{B}_{a\rightarrow i}^{(\text{d})} &=& \beta^2 J^2_a ~\left[ \prod_{k\in \partial a \setminus i} h^2_{k\rightarrow a} -
  \prod_{k\in \partial a \setminus i} \left(\Delta_{k\rightarrow a}^{(1)} + \Delta_{k\rightarrow a}^{(0)} + h_{k\rightarrow a}^2 \right)\right] \nonumber \\
\hat{B}_{a\rightarrow i}^{(\text{nd})} &=& \beta^2 J^2_a~\left[ \prod_{k\in \partial a \setminus i} \left(\Delta_{k\rightarrow a}^{(0)} + h_{k\rightarrow a}^2\right)
  -\prod_{k\in \partial a \setminus i} h^2_{k\rightarrow a}\right]. \nonumber \\
\label{eq:multiv-gauss-coeffs}
\eea
The \emph{function-to-variable} message, expressed by
Eq.~(\ref{eq:cavity2-1RSB}), reads therefore as
\be
\hat{\eta}_{a\rightarrow i}(\boldsymbol{\sigma}_i) \propto
\exp\left\lbrace \hat{A}_{a\rightarrow i}~\sum_{\alpha=1}^x \sigma_i^\alpha -\frac{1}{2}~\left( \hat{B}_{a\rightarrow i}^{(\text{d})} + \frac{\lambda}{K} \right)~\sum_{\alpha=1}^x (\sigma_i^\alpha)^2 +
\frac{1}{2}~\hat{B}_{a\rightarrow i}^{(\text{nd})}~ \sum_{\alpha\neq\beta}^x \sigma_i^\alpha\sigma_i^\beta\right\rbrace,
\ee
while from Eq.~(\ref{eq:cavity1-1RSB}) we have that the
\emph{variable-to-function} message reads as
\be
\eta_{i\rightarrow a}(\boldsymbol{\sigma}_i) \propto \exp\left\lbrace \left(\sum_{b\in \partial i \setminus a} \hat{A}_{b\rightarrow i}\right)~\sum_{\alpha=1}^x \sigma_i^\alpha -\frac{1}{2}~
\left( \sum_{b\in \partial i \setminus a} \hat{B}_{b\rightarrow i}^{(\text{d})} + \lambda \right)~\sum_{\alpha=1}^x (\sigma_i^\alpha)^2 +
\frac{1}{2}~\left( \sum_{b\in \partial i \setminus a} \hat{B}_{b\rightarrow i}^{(\text{nd})}\right)~ \sum_{\alpha\neq\beta}^x \sigma_i^\alpha\sigma_i^\beta\right\rbrace.
\label{eq:marginal-new}
\ee
In order to keep the notation simple let us define:
\bea
A_{i\rightarrow a} &\equiv& \sum_{b\in \partial i \setminus a}\hat{A}_{b\rightarrow i} = \sum_{b\in \partial i \setminus a} \beta J_b \prod_{k\in \partial b \setminus i} h_{k\rightarrow b} \nonumber \\
B^{(\text{d})}_{i\rightarrow a} &\equiv& \lambda + \sum_{b\in \partial i \setminus a}\hat{B}_{a\rightarrow i}^{(\text{d})} =\lambda - \sum_{b\in \partial i \setminus a} \beta^2 J^2_b ~\left[
  \prod_{k\in \partial b \setminus i} \left(\Delta_{k\rightarrow b}^{(1)} + \Delta_{k\rightarrow b}^{(0)} + h_{k\rightarrow b}^2 \right) - \prod_{k\in \partial b \setminus i} h^2_{k\rightarrow b}\right] \nonumber \\
B^{(\text{nd})}_{i\rightarrow a} &\equiv& \sum_{b\in \partial i \setminus a} \hat{B}_{a\rightarrow i}^{(\text{nd})}= \sum_{b\in \partial i \setminus a}  \beta^2 J^2_b~\left[ \prod_{k\in \partial b \setminus i} \left(\Delta_{k\rightarrow b}^{(0)} + h_{k\rightarrow b}^2\right) - \prod_{k\in \partial b \setminus i} h_{k\rightarrow b}^2 \right]\nonumber, \\
\eea
so that Eq. (\ref{eq:cavity1-1RSB}) can be rewritten in the more compact form as:
\be
\eta_{i\rightarrow a}(\boldsymbol{\sigma}_i) \propto \exp\left\lbrace A_{i\rightarrow a}\sum_{\alpha=1}^x \sigma_i^\alpha -\frac{1}{2}B^{(\text{d})}_{i\rightarrow a}\sum_{\alpha=1}^x (\sigma_i^\alpha)^2 +
\frac{1}{2}B^{(\text{nd})}_{i\rightarrow a} \sum_{\alpha\neq\beta}^x \sigma_i^\alpha\sigma_i^\beta\right\rbrace.
\label{eq:marginal-new-compact}
\ee
The expression above can be further simplified by introducing the matrix $\mM_{\alpha\beta}$ and the vector $u_\alpha$ such that
\bea
\nonumber 
u_\alpha &=& \frac{A_{i\rightarrow a}}{B^{(\text{d})}_{i\rightarrow a}-(x-1)B^{(\text{nd})}_{i\rightarrow a}} ~~~~ \forall\alpha,\\
\mM_{\alpha\beta} &=& \delta_{\alpha\beta}~B^{(\text{d})}_{i\rightarrow a} + (1-\delta_{\alpha\beta})~(-B^{(\text{nd})}_{i\rightarrow a}) 
\label{eq:matrixM}
\eea
and the normalized distribution, written in the standard form for a multivariate Gaussian, reads
\be
\eta_{i\rightarrow a}(\boldsymbol{\sigma}_i) =  \sqrt{\frac{\det \mM}{(2\pi)^x}} \exp\left\lbrace -\frac{1}{2} (\boldsymbol{\sigma}_i-\bf u)^T \mM  (\boldsymbol{\sigma}_i-\bf u) \right\rbrace,
\ee
The closed cavity equations, which in the 1RSB case are three rather than two, are simply obtained
by taking the averages in Eq.~(\ref{eq:1RSB-parameters})
with respect to the marginal distribution $\eta_{i\rightarrow a}(\boldsymbol{\sigma}_i)$:
\bea
h_{i\rightarrow a} &=& \langle \sigma_i^\alpha \rangle = u_\alpha \nonumber  \\
\Delta^{(0)}_{i\rightarrow a} &=& \langle \sigma_i^\alpha \sigma_i^\beta \rangle - [\langle \sigma_i^\alpha \rangle]^2 = \mM^{-1}_{\alpha\beta} \nonumber \\
\Delta^{(1)}_{i\rightarrow a} &=& \langle [\sigma_i^\alpha]^2 \rangle  - \langle \sigma_i^\alpha \sigma_i^\beta \rangle = \mM^{-1}_{\alpha\alpha}-\mM^{-1}_{\alpha\beta},
\nonumber
\eea
where the general expression of the inverse matrix element is:
\be
\mM^{-1}_{\alpha\beta} = \frac{1}{\left(B^{(\text{d})}_{i\rightarrow a}+B^{(\text{nd})}_{i\rightarrow a}\right)}~\delta_{\alpha\beta} +
\frac{B^{(\text{nd})}_{i\rightarrow a}}{\left(B^{(\text{d})}_{i\rightarrow a}+B^{(\text{nd})}_{i\rightarrow a}\right)\left(B^{(\text{d})}_{i\rightarrow a}+(1-x)B^{(\text{nd})}_{i\rightarrow a}\right)}.
\label{eq:inverse-M}
\ee
For the ease of the reader willing to implement them in a code, let us
write explicitly the closed cavity equations:
\bea
h_{i\rightarrow a} &=& \frac{\beta \sum_{b\in \partial i\setminus a}J_b \prod_{k\in\partial b\setminus i} h_{k\rightarrow b}  }
{\mD^{(1)}_{i\to a} - x \mD^{(0)}_{i\to a}
\nonumber
   }  \\
\Delta^{(0)}_{i\rightarrow a} &=& \frac{
\mD^{(0)}_{i\to a} }{
 \mD^{(1)}_{i\to a} \left(\mD^{(1)}_{i\to a} - x \mD^{(0)}_{i\to a} \right)}
\label{eq:belief-prop}
    \\
\Delta^{(1)}_{i\rightarrow a} &=& \frac{1}{\mD^{(1)}_{i\to a}}
\nonumber
\eea
with
\bea
\mD^{(1)}_{i\to a} &\equiv&  \lambda - \beta^2\sum_{b\in\partial i\setminus a}J_b^2\left[  \prod_{k\in\partial b\setminus i}\left(\Delta^{(0)}_{k\rightarrow b}+\Delta^{(1)}_{k\rightarrow b}+h^2_{k\rightarrow b}\right) - \prod_{k\in\partial b\setminus i}\left(\Delta^{(0)}_{k\rightarrow b}+h^2_{k\rightarrow b}\right) \right] 
\nonumber
\\
\mD^{(0)}_{i\to a} &\equiv&  \beta^2\sum_{b\in\partial i\setminus a}J_b^2\left[ \prod_{k\in\partial b\setminus i}\left(\Delta^{(0)}_{k\rightarrow b}+h^2_{k\rightarrow b}\right) - 
  \prod_{k\in\partial b\setminus i} h^2_{k\rightarrow b}\right] 
\nonumber
 \eea
While the parameter $\lambda$ is fixed by the normalization condition,
the parameter $x$ is a variational one and has to be choosen in order
to extremize the free-energy, a quantity that is computed
explicitly in the next subsection in the case of a complete graph.

\gblue{In the limit of large mean degree the above saddle point equations can be further simplified by noticing that both $\mD^{(0)}_{i\to a}$ and $\mD^{(1)}_{i\to a}$ concentrate to their mean values, which we denote as $\langle \mD^{(0)}_{i\to a} \rangle = \mD^{(0)}$ and $\langle \mD^{(1)}_{i\to a} \rangle = \mD^{(1)}$, due to the law of large numbers, while $h_{i\to a}$ becomes a Gaussian variable and it is enough to consider its first two moments.
\bea
m \equiv \langle h_{i\rightarrow a} \rangle &=& \frac{\beta \langle J \rangle K ~m^{p-1}}{\mD^{(1)} - x \mD^{(0)}}  \\
q_0 \equiv \langle h_{i\rightarrow a}^2 \rangle &=& \frac{\beta^2 \left[ \langle J^2 \rangle K~q_0^{p-1}+\langle J\rangle^2 K^2 m^{2(p-1)}\right]}{\left(\mD^{(1)} - x \mD^{(0)}\right)^2}  \\
\Delta^{(0)} &=& \frac{\mD^{(0)}}
{\mD^{(1)}\left(\mD^{(1)} - x \mD^{(0)}\right)} \\
\Delta^{(1)} &=& \frac{1}{\mD^{(1)}}
\nonumber
\eea
with
\bea
\mD^{(1)} &\equiv& \lambda - \beta^2 \langle J^2 \rangle K \left[
\left(q_0 + \Delta^{(0)} + \Delta^{(1)}\right)^{p-1} - \left(q_0 + \Delta^{(0)}\right)^{p-1} \right] 
\nonumber
\\
\mD^{(0)} &\equiv& \beta^2 \langle J^2 \rangle K \left[
\left(q_0 + \Delta^{(0)}\right)^{p-1} - q_0^{p-1} \right] 
\nonumber
\eea
and where the symbols $\Delta^{(1)}$ and $\Delta^{(0)}$ represent, respectively,  $\Delta^{(1)} = \langle \Delta^{(1)}_{i\rightarrow a}\rangle$ and $\Delta^{(0)} = \langle \Delta^{(0)}_{i\rightarrow a}\rangle$. By considering the most common model with Gaussian couplings of zero mean ($\langle J \rangle = 0$ and $\langle J^2 \rangle = p!/(2N^{p-1})$) and recalling that $\Delta^{(0)} = q_1-q_0 $, $\Delta^{(1)} = 1-q_1$ and 
\be 
K = \binom{N}{p-1} \sim \frac{N^{p-1}}{(p-1)!}, ~~~\Longrightarrow~~~ \langle J^2 \rangle K \sim \frac{p}{2}
\ee
we are left with the following three closed equations:
\bea
\beta^2 \frac{p}{2} ~q_0^{p-2} &=& \left[ \lambda -p\frac{\beta^2}{2} \left(1-x q_0^{p-1}+(x-1) q_1^{p-1}\right)\right]^2   \\
q_1-q_0 &=& \frac{p\frac{\beta^2}{2}\left(q_1^{p-1}-q_0^{p-1}\right)}{\left[\lambda-p\frac{\beta^2}{2}\left(1-q_1^{p-1}\right)\right]
  \left[ \lambda -p\frac{\beta^2}{2} \left(1-x q_0^{p-1}+(x-1) q_1^{p-1}\right)\right]} \\
1-q_1 &=& \frac{1}{\lambda - p\frac{\beta^2}{2}\left( 1 - q_1^{p-1} \right)} \label{eq:saddle-point-overlaps}
\eea

Let us notice that Eq.~(\ref{eq:saddle-point-overlaps}) allows us to
easily re-express the spherical constraint parameter $\lambda$ as a
function of $q_1$ and $\beta$, i.e.
\be
\lambda = \frac{1}{1-q_1} + p\frac{\beta^2}{2} \left( 1 - q_1^{p-1}\right),
\ee
which will be useful later on. Comparing the expression of the
Langrange multiplier with the one obtained in the RS case,
Eq. (\ref{eq:Lagrange-multiplier}), we see that they are the same with
the substitution $q\to q_1$: in the 1RSB phase the Lagrange multiplier
is enforcing the spherical constraint inside each pure state.

}

\subsection{1RSB free energy}
\label{sec:1RSB-fe}

We now want to
compute the free-energy $\mF(x)$ in the presence of a one-step replica symmetric
Ansatz. The free-energy for the replicated system is \cite{Mezard09,Zamponi}:
\be
\Phi(x) = -\left (\sum_{ a=1}^M \mathbb{F}_a^{
  ^{\text{RSB}}}(x)+ \sum_{i=1}^N \mathbb{F}_i^{ ^{\text{RSB}}}(x) -
\sum_{(ai) \in E} \mathbb{F}_{ai}^{ ^{\text{RSB}}}(x)\right).
\ee
The total free energy of the system is just the free energy of the $x$
coupled replicas, divided by $x$ (and extremized over
$x$).
In principle the three
contributions, respectively representing the {\it
  energetic} and the {\it entropic} contributions and a
normalization, read as:
\be
\beta\mathbb{F}_a^{ ^{\text{RSB}}}(x) = \log\left\lbrace \int \prod_{i\in \partial a} \left[ dm_{i\rightarrow a} Q_{i\to a}(m_{i\rightarrow a})\right]~e^{x \beta\mathbb{F}_a(\lbrace m_{i\rightarrow a} \rbrace)}\right\rbrace
\label{eq:energy-1rsb}
\ee
\be
\beta\mathbb{F}_i^{ ^{\text{RSB}}}(x) = \log\left\lbrace \int \prod_{a \in \partial i} \left[ d\hat{m}_{a \rightarrow i} \hat{Q}_{a\to i}(\hat{m}_{a \rightarrow i})\right]~e^{x \beta\mathbb{F}_i(\lbrace \hat{m}_{a \rightarrow i} \rbrace)}\right\rbrace
\label{eq:entropy-1rsb}
\ee
\be
\beta\mathbb{F}_{ai}^{ ^{\text{RSB}}}(x) =  \log\left\lbrace \int d\hat{m}_{a\rightarrow i}~d{m}_{i\rightarrow a}~\hat{Q}_{a\to i}(\hat{m}_{a \rightarrow i})~Q_{i\to a}(m_{i\rightarrow a})~e^{x \beta\mathbb{F}_{ai}(m_{i \rightarrow a},\hat{m}_{a\rightarrow i} )}\right\rbrace
\label{eq:normalization-1rsb}
\ee
where $\mathbb{F}_a, \mathbb{F}_i$ and $\mathbb{F}_{ai}$ are the RS free energy parts in Eq. (\ref{eq:free-energy-def}), $Q_{i\to a}(m_{i\rightarrow a})$ is the distribution of the cavity marginals that in the case of the 
dense Gaussian 1RSB ansatz has been defined in  Eq. (\ref{eq:Q-1rsb-ansatz}), Sec. \ref{Sec:1rsbAnsatz}
and the distribution $\hat{Q}_{a\to i}(\hat{m}_{a \rightarrow i})$ should be the one of the function-to-variable fields.

The detailed computation is reported in the Appendix \ref{app:1RSBfree_energy}, here we just write the result that is the same found with the replica approach \cite{Crisanti92}:
\bea
-x\beta\mathcal{F}(x) &=&
\sum_{a=1}^M \beta\mathbb{F}_a^{\text{RSB}} + \sum_{i=1}^N \beta\mathbb{F}_i^{\text{RSB}} -\sum_{(ai)\in E} \beta\mathbb{F}_{ai}^{\text{RSB}} \nonumber \\ 
&=& \frac{xN}{2} \Big\lbrace ~\frac{\beta^2}{2} \left[ 1 - (1-x)q_1^p - x q_0^p\right] + \frac{q_0}{\left[ 1 - xq_0 - (1-x) q_1 \right]} +
\frac{x-1}{x} \log\left( 1-q_1\right)+\frac{1}{x}\log\left[ 1-xq_0-(1-x)q_1\right] \Big\rbrace \nonumber \\
\eea


\section{Conclusions}
\label{sec:4}

In this paper we have derived the cavity equations for solving diluted spherical $p$-spin models. Such a cavity-based derivation makes evident the underlying assumption that reflects itself in the distribution of local fields: in the RS ansatz replicas are uncorrelated and have Gaussian local fields, while in the 1RSB ansatz replicas have correlated Gaussian local fields whose covariance matrix depends on whether the replicas are in the same state or not, as pointed out in Ref. \cite{AKUZ19}.

We have derived the cavity equations exploiting the same high-temperature expansion that leads to mean-field approximations \cite{MFLKMZ19}.

We have solved the cavity equations in the fully-connected case. In this case the solution is homogeneous, depends on very few parameters and can be written explicitly, leading to the same expression that was obtained via the replica method.

The approach based on the cavity method has several advantages:
\begin{compactitem}
    \item it makes clear the underlying assumptions;
    \item it holds also for the diluted version of the model (provided the solution does not condensate);
    \item it can be converted in message-passing algorithms, the RS Belief Propagation and the 1RSB Survey Propagation;
    \item it allows to study heterogeneous solutions in diluted models, until the condensation transition.
\end{compactitem}

Our work, besides
providing the first complete reference on the equivalence between the
replica and the cavity methods for spherical disordered models, paves
the way to a more systematic study of \emph{inhomogeneous} glassy
phases in diluted mean-field models and represents a reference point for the
systematic development of algorithms for combinatorial optimization
and inference problems characterized by continuous
variables~\cite{Marruzzo18,H18,AKUZ19,MLKZ20}. 

\begin{acknowledgments}
The research has been supported by the European Research Council under the European Unions Horizon 2020 research and innovation programme (grant No.~694925, G.~Parisi).
G.~G.~acknowledges the financial support of the Simons Foundation (Grant No.~454949, G.~Parisi) and thanks the Physics Department of Sapienza, University of Rome for kind hospitality in the first stages of this work.
\end{acknowledgments}

\appendix

\section{Replica Symmetric free energy} 
\label{app:RSFreeEnergy}

In this section we compute the replica-symmetric
free energy of the model, which is defined as~\cite{Mezard09}:
\be
-\beta F \equiv \beta\left(\sum_{a=1}^{M} \mathbb{F}_a + \sum_{i=1}^N
 \mathbb{F}_i - \sum_{(ai) \in E} \mathbb{F}_{ai}\right)\equiv\sum_{a=1}^{M} \log(Z_a) + \sum_{i=1}^N
\log(Z_i) - \sum_{(ai) \in E}\log(Z_{(ai)}),
\label{eq:free-energy-def-app}
\ee
where we have respectively
\bea
Z_a &=& \int_{-\infty}^{\infty} \prod_{i\in \partial a} d\sigma_i
~\eta_{i\rightarrow a}(\sigma_i) ~ e^{\beta J_a \prod_{i\in \partial
    a}\sigma_i}  \label{eq:free-energy-Za-app} \\ 
Z_i &=& \int_{-\infty}^{\infty} d\sigma_i ~ \prod_{a\in \partial i}
\hat{\eta}_{a\rightarrow i}(\sigma_i)  \label{eq:free-energy-Zi-app} \\ 
Z_{(ai)} &=&  \int_{-\infty}^{\infty} d\sigma_i ~
\hat{\eta}_{a\rightarrow i}(\sigma_i) ~\eta_{i\rightarrow a}(\sigma_i) \label{eq:free-energy-Zai-app}.
\eea
Let us now compute the three contributions to the free-energy,
starting from the simplest, which is the energy per functional node
$\log(Z_a)$. \gblue{ By expanding the Boltzmann weight in Eq.~(\ref{eq:free-energy-Za-app}) we get:
\bea Z_a &\simeq&\int \left(\prod_{i\in \partial a} d\sigma_i
~\eta_{i\rightarrow a}(\sigma_i)\right)~\left[ 1 + \beta
  J_a\prod_{i\in\partial
    a}\sigma_i+\frac{\beta^2}{2}J_a^2\prod_{i\in\partial a}\sigma_i^2
  \right] \nonumber \\ &=& \left[ 1 + \beta J_a~\prod_{i\in\partial a}
  m_{i\rightarrow a} + \frac{\beta^2}{2}J_a^2 \prod_{i\in\partial a}
  \left( v_{i\rightarrow a} + m_{i\rightarrow a}^2 \right)\right] \nonumber \\ 
&\simeq &  \exp\left\lbrace \beta J_a~\prod_{i\in\partial a}
m_{i\rightarrow a} + \frac{\beta^2}{2}J_a^2
\left(\prod_{i\in\partial a} \left( v_{i\rightarrow a} + m_{i\rightarrow a}^2 \right) -\prod_{i\in\partial a} m^2_{i\rightarrow a}\right)
\right\rbrace.
\label{eq:Za-compute-first}
\eea
We have, therefore, that the sum over all the $M = \binom{N}{p}$ interaction nodes  reads as
\begin{align}
  \sum_{a=1}^{M}\log\left( Z_a\right) &= \beta \sum_{a=1}^{M}J_a~\prod_{i\in\partial a} m_{i\rightarrow a} +
  \frac{\beta^2}{2} \sum_{a=1}^{M} J_a^2 \left( \prod_{i\in\partial a} \left( v_{i\rightarrow a} + m_{i\rightarrow a}^2 \right) - \prod_{i\in\partial a} m_{i\rightarrow a}^2 \right)   \nonumber \\ 
  &= \beta m^p M \langle J \rangle  + \frac{\beta^2}{2} (1-q^p) M \langle J^2 \rangle \nonumber \\
  & = N \frac{\beta^2}{4} (1-q^p) J_2,
\label{eq:Za-compute-final}
\end{align}
since we have assumed $\langle J \rangle = 0$ and for large $N$
\begin{align}
  M \langle J^2\rangle = \frac{M J_2 p!}{2N^{p-1}} \simeq N ~J_2/2.
\end{align}

Let us now compute the entropy per spin $\log(Z_i)$.
From Eq.~(\ref{eq:free-energy-Zi}) we have that:

\bea
&& Z_i = \int_{-\infty}^\infty d\sigma_i \prod_{a\in\partial i} \left[ \frac{1}{\hat{Z}_{a\rightarrow i}}\exp\left\lbrace - \frac{\lambda}{2K}\sigma_i^2 + \beta J_a \left( \prod_{j\in\partial a \setminus i} m_{i\rightarrow a}\right)~\sigma_i + \frac{\beta^2}{2}J_a^2\left(\prod_{j\in\partial a \setminus i} \left( v_{j\rightarrow a} + m_{j\rightarrow a}^2 \right)-\prod_{j\in\partial a \setminus i}  m_{j\rightarrow a}^2\right)~ \sigma_i^2 \right\rbrace\right] \nonumber \\
&=& \left[ \prod_{a\in\partial i} \frac{1}{\hat{Z}_{a\rightarrow i}} \right] \int_{-\infty}^\infty d\sigma_i \exp\left\lbrace - \frac{\lambda}{2}\sigma_i^2 +
\beta \sigma_i \sum_{a=1}^K J_a \left( \prod_{j\in\partial a \setminus i} m_{i\rightarrow a}\right) + \frac{\beta^2}{2} \sigma_i^2 \sum_{a=1}^K  J_a^2\left(\prod_{j\in\partial a \setminus i} \left( v_{j\rightarrow a} + m_{j\rightarrow a}^2 \right)-\prod_{j\in\partial a \setminus i}  m_{j\rightarrow a}^2\right)\right\rbrace. \nonumber \\
\label{eq:Zi-compute-1st}
\eea
Since to the leading order we have
\be
\sum_{a=1}^K  J_a^2\left(\prod_{j\in\partial a \setminus i} \left( v_{j\rightarrow a} + m_{j\rightarrow a}^2 \right)-\prod_{j\in\partial a \setminus i}  m_{j\rightarrow a}^2\right) \simeq \sum_{a\in\partial i \setminus b} J_a^2\left(\prod_{j\in\partial a \setminus i} \left( v_{j\rightarrow a} + m_{j\rightarrow a}^2 \right)-\prod_{j\in\partial a \setminus i}  m_{j\rightarrow a}^2\right), 
\ee
we can take advantage of the cavity equations
[Eq.~(\ref{eq:closure-cavity})] and of the fact that both
$m_{j\rightarrow a}^2$ and $v_{j\rightarrow a}$ are variables which
concentrate in the large $N$ limit to write:
\be
\beta^2 \sum_{a=1}^K  J_a^2\left(\prod_{j\in\partial a \setminus i} \left( v_{j\rightarrow a} + m_{j\rightarrow a}^2 \right)-\prod_{j\in\partial a \setminus i}  m_{j\rightarrow a}^2\right) = \lambda - \frac{1}{1-q},
\ee
which allows us to write
\begin{align}
Z_i &= \left[ \prod_{a\in\partial i}\hat{Z}_{a\rightarrow i}^{-1} \right] \int_{-\infty}^\infty d\sigma_i~\exp\left\lbrace -\frac{1}{2(1-q)}~\sigma_i^2+
\beta~\sigma_i \sum_{a=1}^K J_a \left( \prod_{j\in\partial a \setminus i} m_{j\rightarrow a}\right) \right\rbrace \nonumber \\
& = \left[ \prod_{a\in\partial i}\hat{Z}_{a\rightarrow i}^{-1} \right] \sqrt{2\pi}~(1-q)^{1/2}\exp\left\lbrace \frac{(1-q)}{2} 
\left[ \beta \sum_{a=1}^K J_a \left( \prod_{j\in\partial a \setminus i} m_{j\rightarrow a}\right) \right]^2\right\rbrace.
\label{eq:Zi-compute-2nd}
\end{align}
By taking the square of the last summation left in the argument of the
exponential function in Eq.~(\ref{eq:Zi-compute-2nd}) we get 
\begin{align}
  \beta^2 \left[ \sum_{a=1}^K J_a \prod_{j\in\partial a \setminus i} m_{j\rightarrow a} \right]^2 & =
  \beta^2 \sum_{a=1}^K J_a^2 \prod_{j\in\partial a \setminus i} m^2_{j\rightarrow a} +
  \sum_{a\neq b}^K J_a J_b \left(\prod_{j\in\partial a \setminus i} m_{j\rightarrow a}\right)~\left( \prod_{k\in\partial a \setminus i} m_{k\rightarrow a}\right) \nonumber \\
  & \simeq \beta^2 K \langle J^2\rangle q^{p-1} + K^2 \langle J \rangle^2 m^{2p-2} = \frac{q}{(1-q)^2},
  \label{eq:lin-squared}
\end{align}
where for the rightmost identity in the last line of
Eq.~(\ref{eq:lin-squared}) we used Eq.~(\ref{eq:cavity-av-moments-2})
and the definition of the spherical constraint. By then plugging the result
of Eq.~(\ref{eq:lin-squared}) into Eq.~(\ref{eq:Zi-compute-2nd}) we
get
\be
Z_i = \left[ \prod_{a\in\partial i}\hat{Z}_{a\rightarrow i}^{-1} \right] \sqrt{2\pi}~(1-q)^{1/2}\exp\left\lbrace \frac{1}{2}\frac{q}{(1-q)}\right\rbrace.
\label{eq:Zi-compute-3rd}
\ee
From the expression of Eq.~(\ref{eq:Zi-compute-3rd}), summing over all spins (and neglecting constant terms) one finds:
\be
\sum_{i=1}^N\log(Z_i) = \frac{N}{2}\left[\log(1-q)+\frac{q}{(1-q)}\right]- \sum_{i=1}^N\sum_{a\in\partial i} \log(\hat{Z}_{a\rightarrow i}). 
\label{eq:Zi-compute-final}
\ee
Finally, the contribution to the free-energy coming from the edges reads as:
\begin{align}
& Z_{(ai)} =  \int_{-\infty}^{\infty} d\sigma_i ~
\hat{\eta}_{a\rightarrow i}(\sigma_i) ~\eta_{i\rightarrow a}(\sigma_i) \nonumber \\
& = \int_{-\infty}^{\infty} d\sigma_i ~\frac{\exp\left\lbrace -\frac{(\sigma_i-m_{i\rightarrow a})^2}{2v_{i\rightarrow a}} -
\frac{\lambda}{2K} ~ \sigma_i^2 + \beta  J_a 
\sigma_i \prod_{j\in\partial a \setminus i} m_{j\rightarrow a} +
\frac{1}{2} \beta^2  J_a^2 \left( \prod_{j\in\partial a \setminus i} (v_{j\rightarrow a}+m_{j\rightarrow a}^2)
- \prod_{j\in\partial a \setminus i} m^2_{j\rightarrow a} \right) ~\sigma_i^2 \right\rbrace}{\hat{Z}_{a\rightarrow i}\sqrt{2 \pi v_{i\rightarrow a}} } \nonumber \\
\label{eq:Zai-compute-first}
\end{align}
Since in the large-$N$ limit we have, respectively,
\begin{align}
  \frac{(\sigma_i-m_{i\rightarrow a})^2}{2v_{i\rightarrow a}}  &= \mO(1) \nonumber \\
  |J_a| & = \mO\left(\frac{1}{N^{(p-1)/2}}\right) \nonumber \\
  J_a^2 & = \mO\left(\frac{1}{N^{p-1}}\right) \nonumber \\
  \frac{1}{K} &= \mO\left(\frac{1}{N^{p-1}}\right),
\label{eq:Zai-compute-2nd}
\end{align}
all terms added inside the argument of the exponential
Eq.~(\ref{eq:Zai-compute-first}) are subleading with respect to the
first one, so that one can write:
\be
Z_{(ai)} \simeq \frac{1}{\hat{Z}_{a\rightarrow i}} \int_{-\infty}^{\infty} \frac{d\sigma_i}{\sqrt{2 \pi v_{i\rightarrow a}}}
~e^{-\frac{(\sigma_i-m_{i\rightarrow a})^2}{2v_{i\rightarrow a}} } = \frac{1}{\hat{Z}_{a\rightarrow i}}
\label{eq:Zai-compute-3rd}
\ee
Finally, summing over all the edges we get:
\be
\sum_{(ai)\in E} \log(Z_{(ai)}) = - \sum_{(ai)\in E}\log(\hat{Z}_{a\rightarrow i}) = -\sum_{i=1}^N\sum_{a\in\partial i}\log(\hat{Z}_{a\rightarrow i}),
\label{eq:Zai-compute-final}
\ee
which, in turn, allows us to write
\begin{align}
  \sum_{i = 1}^N \log(Z_i) - \sum_{(ai)\in E} \log(Z_{(ai)}) & = \frac{N}{2}\left[\log(1-q)+\frac{q}{(1-q)}\right]- \sum_{i=1}^N\sum_{a\in\partial i} \log(\hat{Z}_{a\rightarrow i}) + \sum_{i=1}^N\sum_{a\in\partial i}\log(\hat{Z}_{a\rightarrow i}) \nonumber \\
  & = \frac{N}{2}\left[\log(1-q)+\frac{q}{(1-q)}\right].
\end{align}  

By finally taking into account also the contribution of $\sum_{a=1}^M
\log(Z_a)$ we get:
\bea
-\beta F_{\textrm{RS}} &=&
\frac{N}{2}\left[\frac{\beta^2}{2} (1-q^p) J_2 + \log(1-q)+ \frac{q}{(1-q)}\right],
\label{eq:frs-final-app}
\eea
where we used the fact that $N\sum_{a\in\partial i} = \sum_{(ai)\in
  E}$.}\\

\section{One-step Replica Symmetry Breaking solution}
\label{app:1RSB}

\subsection{1RSB Cavity Equations}
\label{app:cavity-eq}

We derive here some useful identities which can be obtained from the
1RSB cavity equations.  One has just to equate the coefficients of
the diagonal, i.e. $\sum_{\alpha=1}^x (\sigma^\alpha)^2$, and
off-diagonal, i.e. $\sum_{\alpha\neq\beta}^x \sigma^\alpha\sigma^\beta
$ on the left and on the right of the first of the cavity equations in
Eq.~(\ref{eq:cavity2-1RSB}). Before doing this let us just write the
local 1RSB marginal $\eta(\boldsymbol{\sigma}_i)$ explicitly, i.e. we
compute explicitly the integral over $m_{i\rightarrow a}$ in
Eq.~(\ref{eq:1RSB-ansatz}):
\bea
&& \eta_{i\rightarrow a}(\boldsymbol{\sigma}_i) = \int_{-\infty}^\infty dm_{i\rightarrow a}~\frac{1}{\sqrt{2\pi\Delta_{i\rightarrow a}^{(0)}}}
\exp\left( -\frac{(m_{i\rightarrow a}-h_{i\rightarrow a})^2}{2\Delta_{i\rightarrow a}^{(0)}}\right)
\frac{1}{\left[ \sqrt{2\pi\Delta_{i\rightarrow a}^{(1)}}\right]^x} \exp\left(-\sum_{\alpha=1}^x\frac{(\sigma_i^\alpha-m_{i\rightarrow a})^2}{2\Delta_{i\rightarrow a}^{(1)}} \right) \nonumber \\
&\approx& \exp\left\lbrace-\frac{1}{2\Delta^{(1)}_{i\rightarrow a}}\sum_{\alpha=1}^x \sigma_\alpha^2 - \frac{h_{i\rightarrow a}}{2\Delta^{(0)}_{i\rightarrow a}}\right\rbrace
\int_{-\infty}^\infty~dm_{i\rightarrow a}~\exp\left\lbrace -\frac{1}{2}\left( \frac{\Delta^{(1)}_{i\rightarrow a}+x\Delta^{(0)}_{i\rightarrow a}}{\Delta^{(1)}_{i\rightarrow a}\Delta^{(0)}_{i\rightarrow a}}\right)
m^2_{i\rightarrow a} - \left( \frac{h_{i\rightarrow a}}{\Delta^{(0)}_{i\rightarrow a}}+\frac{1}{\Delta^{(1)}}\sum_{\alpha=1}^x \sigma^\alpha\right) m_{i\rightarrow a} \right\rbrace \nonumber \\ 
&\approx&
\exp\left\lbrace-\frac{1}{2\Delta^{(1)}_{i\rightarrow a}}\sum_{\alpha=1}^x \sigma_\alpha^2 - \frac{h_{i\rightarrow a}}{2\Delta^{(0)}_{i\rightarrow a}} +
\frac{1}{2} \left( \frac{\Delta^{(1)}_{i\rightarrow a}\Delta^{(0)}_{i\rightarrow a}}{\Delta^{(1)}_{i\rightarrow a}+x\Delta^{(0)}_{i\rightarrow a}} \right)
\left( \frac{h_{i\rightarrow a}}{\Delta^{(0)}_{i\rightarrow a}}+\frac{1}{\Delta^{(1)}_{i\rightarrow a}}\sum_{\alpha=1}^x \sigma^\alpha\right)^2 \right\rbrace \nonumber \\
&\approx& \exp\left\lbrace -\frac{1}{2\Delta^{(1)}_{i\rightarrow a}} \left( 1-\frac{\Delta^{(0)}_{i\rightarrow a}}{\Delta^{(1)}_{i\rightarrow a}+x\Delta^{(0)}_{i\rightarrow a}}\right) \sum_{\alpha=1}^x \sigma_\alpha^2
+ \frac{1}{2\Delta^{(1)}_{i\rightarrow a}} \left(\frac{\Delta^{(0)}_{i\rightarrow a}}{\Delta^{(1)}_{i\rightarrow a}+x\Delta^{(0)}_{i\rightarrow a}}\right) \sum_{\alpha\neq\beta}^x \sigma_\alpha\sigma_\beta  +
\left(\frac{h_{i\rightarrow a}}{\Delta^{(1)}_{i\rightarrow a}+x\Delta^{(0)}_{i\rightarrow a}}\right) \sum_{\alpha=1}^x \sigma_\alpha \right\rbrace, \nonumber \\
\label{eq:cavity-1RSB-left}
\eea
where, in the last line of Eq.~(\ref{eq:cavity-1RSB-left}), we have
retained only terms at least linear in $\sigma_i^\alpha$.  Thus,
matching the coefficients of the linear and quadratic in
$\sigma_i^\alpha$, both diagonal and non-diagonal, appearing in the
expression of $\eta({\bf \sigma}^\alpha_i)$ in
Eq.~(\ref{eq:cavity-1RSB-left}) above here and in
Eq.~(\ref{eq:marginal-new-compact}) we get the set of three equations:
\bea
 \frac{h_{i\rightarrow a}}{\Delta^{(1)}_{i\rightarrow a}+x\Delta^{(0)}_{i\rightarrow a}} &=& \beta \sum_{b\in \partial i \setminus a} J_b \prod_{k\in \partial b \setminus i} h_{k\rightarrow b} = A \nonumber \\
 \frac{1}{\Delta^{(1)}_{i\rightarrow a}} \left( 1-\frac{\Delta^{(0)}_{i\rightarrow a}}{\Delta^{(1)}_{i\rightarrow a}+x\Delta^{(0)}_{i\rightarrow a}}\right) &=&
\lambda + \sum_{b\in \partial i \setminus a} \beta^2 J^2_b ~\left[ \prod_{k\in \partial b \setminus i} h^2_{k\rightarrow b} -
  \prod_{k\in \partial b \setminus i} \left(\Delta_{k\rightarrow b}^{(1)} + \Delta_{k\rightarrow b}^{(0)} + h_{k\rightarrow b}^2 \right)\right]
= B^{(\text{d})}\nonumber \\ \nonumber \\
  \frac{1}{\Delta^{(1)}_{i\rightarrow a}} \left(\frac{\Delta^{(0)}_{i\rightarrow a}}{\Delta^{(1)}_{i\rightarrow a}+x\Delta^{(0)}_{i\rightarrow a}}\right) &=&
\sum_{b\in \partial i \setminus a}  \beta^2 J^2_b~\left[ \prod_{k\in \partial b \setminus i} \left(\Delta_{k\rightarrow b}^{(0)} + h_{k\rightarrow b}^2\right) - \prod_{k\in \partial b \setminus i} h_{k\rightarrow b}^2 \right] = B^{(\text{nd})}\nonumber \\ 
\label{eq:variable-function-equations}
\eea

\gblue{ Since we known that the three variables
  $\Delta^{(1)}_{k\rightarrow b}$, $\Delta^{(0)}_{k\rightarrow b}$ and
  $h^2_{k\rightarrow b}$ all have a distribution which become
  concentrated in the large $N$ limit, we have that the summations on
  the right hand side of the last two lines in
  Eq.~(\ref{eq:variable-function-equations}) read as
  \begin{align}
     \lambda + \sum_{b\in \partial i \setminus a} \beta^2 J^2_b ~\left[ \prod_{k\in \partial b \setminus i} h^2_{k\rightarrow b} -
      \prod_{k\in \partial b \setminus i} \left(\Delta_{k\rightarrow b}^{(1)} + \Delta_{k\rightarrow b}^{(0)} + h_{k\rightarrow b}^2 \right)\right] &=
    \lambda - \beta^2 K \langle J^2 \rangle (1-q_0^{p-1}) = B^{\text{(d)}}\nonumber \\
     \sum_{b\in \partial i \setminus a}  \beta^2 J^2_b~\left[ \prod_{k\in \partial b \setminus i} \left(\Delta_{k\rightarrow b}^{(0)} + h_{k\rightarrow b}^2\right) - \prod_{k\in \partial b \setminus i} h_{k\rightarrow b}^2 \right] &= \beta^2 K \langle J^2 \rangle (q_1^{p-1}-q_0^{p-1}) = B^{\text{(nd)}},
     \label{eq:variable-function-equations-2}
    \end{align}

  where we have introduced the symbols $B^{\text{(d)}}$ and
  $B^{\text{(nd)}}$ to highlight that dependence on the subscript
  indices can be dropped. Thus, since expressions on the \emph{right}
  hand side of the last two lines of
  Eq.~(\ref{eq:variable-function-equations}) are summations over a
  very large number of local fields, as a consequence one can drop the
  site index for the expressions on the \emph{left} hand side. This
  means that inside the left hand members in
  Eq.~(\ref{eq:variable-function-equations}), except the equation in
  the first line, the fields $\Delta^{(1)}_{i\rightarrow a}$,
  $\Delta^{(0)}_{i\rightarrow a}$ and $h^2_{i\rightarrow a}$ can be
  replaced everywhere with their average values,
  \bea
  \langle h^2_{i\rightarrow a} \rangle &=& q_0 \nonumber \\ \langle \Delta^{(0)}_{i\rightarrow
    a} \rangle &=& q_1-q_0 \nonumber \\ \langle \Delta^{(1)}_{i\rightarrow a} \rangle &=&
  1-q_1,
  \eea
  leading finally to the following expressions
  \begin{align}
    B^{(\text{d})} &= \frac{1}{1-q_1} \left[ 1-\frac{q_1-q_0}{1-xq_0-(1-x)q_1}\right] \nonumber \\
    B^{(\text{nd})} &= \frac{q_1-q_0}{1-q_1} \cdot \frac{1}{1-xq_0-(1-x)q_1}.
    \label{eq:1RSB-matrix-elements}
  \end{align}

  For what concerns the first line of
  Eq.~(\ref{eq:variable-function-equations}), while it is in general
  not possible to assume concentration for the distribution of
  $h_{i\rightarrow a}$, this can be done for the distribution of
  $h_{i\rightarrow a}^2$. This fact can be exploited to write useful
  relation for the squared variable:

  \begin{align}
    A_{i\rightarrow a}^2 &= \beta^2 \left[\sum_{b\in \partial i \setminus a} J_b \prod_{k\in \partial b \setminus i} h_{k\rightarrow b}\right]^2 \nonumber \\
    & = \beta^2 \sum_{b\in \partial i \setminus a} J_b^2 \prod_{k\in \partial b \setminus i} h^2_{k\rightarrow b} +
    \beta^2 \sum_{b\neq c \in \partial i \setminus a} J_b J_a \prod_{k\in \partial b \setminus i} h_{k\rightarrow b} \prod_{j\in \partial c \setminus i} h_{j\rightarrow c} \nonumber \\
    & \simeq \beta^2 \langle J^2 \rangle \langle h^{2(p-1)} \rangle + \langle J \rangle^2 \langle h \rangle^{2(p-1)},
    \label{eq:Asquared-1}
  \end{align}
  which tells us that we can also write
  \begin{align}
    A_{i\rightarrow a}^2 = \left\langle \frac{h^2_{i\rightarrow a}}{\left(\Delta^{(1)}_{i\rightarrow a}+x\Delta^{(0)}_{i\rightarrow a}\right)^2} \right\rangle =\frac{\langle h^2_{i\rightarrow a}\rangle}{\left(\langle \Delta^{(1)}_{i\rightarrow a}\rangle+x\langle \Delta^{(0)}_{i\rightarrow a}\rangle \right)^2} = 
    \frac{q_0}{\left( 1- x q_0 - (1-x) q_1\right)^2}.
    \label{eq:Asquared-2}
  \end{align}

}

\subsection{1RSB Free energy}
\label{app:1RSBfree_energy}

Let us reproduce here the exact formulas that we need to
compute $\mF(x)$ in the presence of a one-step replica symmetry
Ansatz. The free-energy for the replicated system is \cite{Mezard09,Zamponi}:
\be
\Phi(x) = -\left (\sum_{ a=1}^M \mathbb{F}_a^{
  ^{\text{RSB}}}(x)+ \sum_{i=1}^N \mathbb{F}_i^{ ^{\text{RSB}}}(x) -
\sum_{(ai) \in E} \mathbb{F}_{ai}^{ ^{\text{RSB}}}(x)\right).
\ee
In principle the three
contributions, respectively representing the {\it
  energetic} and the {\it entropic} contributions and a
normalization, read as:
\be
\beta\mathbb{F}_a^{ ^{\text{RSB}}}(x) = \log\left\lbrace \int \prod_{i\in \partial a} \left[ dm_{i\rightarrow a} Q_{i\to a}(m_{i\rightarrow a})\right]~e^{x \beta\mathbb{F}_a(\lbrace m_{i\rightarrow a} \rbrace)}\right\rbrace
\label{eq:energy-1rsb-app}
\ee
\be
\beta\mathbb{F}_i^{ ^{\text{RSB}}}(x) = \log\left\lbrace \int \prod_{a \in \partial i} \left[ d\hat{m}_{a \rightarrow i} \hat{Q}_{a\to i}(\hat{m}_{a \rightarrow i})\right]~e^{x \beta\mathbb{F}_i(\lbrace \hat{m}_{a \rightarrow i} \rbrace)}\right\rbrace
\label{eq:entropy-1rsb-app}
\ee
\be
\beta\mathbb{F}_{ai}^{ ^{\text{RSB}}}(x) =  \log\left\lbrace \int d\hat{m}_{a\rightarrow i}~d{m}_{i\rightarrow a}~\hat{Q}_{a\to i}(\hat{m}_{a \rightarrow i})~Q_{i\to a}(m_{i\rightarrow a})~e^{x \beta\mathbb{F}_{ai}(m_{i \rightarrow a},\hat{m}_{a\rightarrow i} )}\right\rbrace
\label{eq:normalization-1rsb-app}
\ee
where $\mathbb{F}_a, \mathbb{F}_i$ and $\mathbb{F}_{ai}$ are the RS free energy parts in Eq. (\ref{eq:free-energy-def}), $Q_{i\to a}(m_{i\rightarrow a})$ is the distribution of the cavity marginals that in the case of the 
dense Gaussian 1RSB ansatz has been defined in  Eq. (\ref{eq:Q-1rsb-ansatz}), Sec. \ref{Sec:1rsbAnsatz}
and the distribution $\hat{Q}_{a\to i}(\hat{m}_{a \rightarrow i})$ should be the one of the function-to-variable fields.
Since we are going to write everything in terms of the {\it
  ``variable to function''} fields $m_{i\rightarrow a}$, it is convenient
to write the probability distribution
$\hat{Q}_{a\to i}(\hat{m}_{a \rightarrow i})$ of {\it ``function to
  variable''} fields $\hat{m}_{ a\rightarrow i}$ in terms of
$Q_{i\to a}(m_{i\rightarrow a})$ and $m_{i\rightarrow a}$. This can be
done taking advantage of the identity \cite{Mezard09}:
\be
\hat{Q}_{a\to i}(\hat{m}_{a \rightarrow i}) = \int \prod_{k\in\partial a \backslash i} \left[ dm_{k\rightarrow a} Q_{k\to a}(m_{k \rightarrow a})\right]~\hat{Z}_{a\to i}^x(\lbrace m_{k\rightarrow a}\rbrace)
\delta\left[ \hat{m}_{ a\rightarrow i} - f(\lbrace m_{k\rightarrow a}\rbrace)\right],
\label{eq:Qhat-def}
\ee
where $f(x)$ is a shorthand to refer to the cavity equations and $\hat{Z}_{a\to i}$ is the normalization factor introduced in the RS equation (\ref{eq:cavity2}). 
By exploiting the last identity  we can
rewrite the expression of the 1RSB entropy per site as 
\be
\beta\mathbb{F}_i^{ ^{\text{RSB}}}(x) = \log \left\lbrace\int \prod_{ a\in\partial i}\prod_{k\in\partial a\backslash i}dm_{k\rightarrow a}~Q_{k\to a}(m_{k \rightarrow a})
~\hat{Z}_{a\to i}^x(\lbrace m_{k\rightarrow a}\rbrace_{k\in\partial a \setminus i})
e^{x \beta\mathbb{F}_i\left(\lbrace f(\lbrace m_{k\rightarrow a}\rbrace) \rbrace_{ a\in\partial i}\right)}
\right\rbrace
\label{eq:entropy-1rsb-Qm}
\ee
 
\subsection{1RSB Energy}
\label{sec:3C}
We compute in what follows the expression of the \emph{energetic} part
of the free-energy.  By plugging into the same expression the
definition of the energetic contribution to the local free energy in
the RS case, see Eq.~(\ref{eq:free-energy-Za}), and the 1RSB weighted
average over the local messages written in Eq.~(\ref{eq:energy-1rsb})
we get:
\bea
\beta\mathbb{F}_a^{ ^{\text{RSB}}}(x) &=& \log\left\lbrace \int \prod_{i\in \partial\alpha} \left[ dm_{i\rightarrow a} Q_{i\to a}(m_{i\rightarrow a})\right]~
\left( \int_{-\infty}^{\infty} \prod_{i\in \partial a} d\sigma_i ~\eta^{\rm s}_{i\rightarrow a}(\sigma_i) ~ e^{\beta J_a \prod_{i\in \partial a}\sigma_i} \right)^x\right\rbrace = \nonumber \\
&=& \log\left\lbrace \int \prod_{i\in \partial\alpha} \left[ dm_{i\rightarrow a} Q_{i\to a}(m_{i\rightarrow a})\right]~
\int \prod_{i\in \partial a} \prod_{\alpha=1}^x d\sigma_i^\alpha ~\eta^{\rm s}_{i\rightarrow a}(\sigma_i^\alpha) ~ e^{\beta J_a \sum_{\alpha=1}^x \prod_{i\in \partial a}\sigma_i^\alpha} \right\rbrace  
\label{eq:free-energy-Za-1RSB}
\eea
The expression in Eq.~(\ref{eq:free-energy-Za-1RSB}) is formally
correct but completely useless unless an explicit ansatz for the
distribution $Q_{i\to a}(m_{i\rightarrow a})$ is given. We consider the
ansatz in Eqs.~(\ref{eq:Q-1rsb-ansatz},\ref{eq:replicas-1rsb-ansatz}), which assumes three parameters, the local magnetization $h_{i\rightarrow a}$
and the two parameters $\Delta_{i\rightarrow a}^{(0)}$ and $\Delta_{i\rightarrow a}^{(1)}$ relative to the coupling
between replicas: 
\bea
\int dm_{i\rightarrow a} Q_{i\to a}(m_{i\rightarrow a}) \prod_{\alpha=1}^x \eta^{\rm s}_{i\rightarrow a}(\sigma_i^\alpha) =
\int_{-\infty}^\infty dm_{i\rightarrow a}~\frac{e^{-(m_{i\rightarrow a}-h_{i\rightarrow a})^2/(2\Delta_{i\rightarrow a}^{(0)})}}{\sqrt{2\pi\Delta_{i\rightarrow a}^{(0)}}}
\prod_{\alpha=1}^x \frac{e^{-(\sigma_i^\alpha-m_{i\rightarrow a})^2/(2\Delta_{i\rightarrow a}^{(1)})}}{\sqrt{2\pi\Delta_{i\rightarrow a}^{(1)}}}, 
\eea
In particular we have that $h_{i\rightarrow a}$ and
$\Delta_{i\rightarrow a}^{(0)}$ are variational parameter for the
probability distribution of the fields $m_{i\rightarrow a}$ on a given
edge, while $\Delta_{i\rightarrow a}^{(1)}$ is a variational parameter
for the coupling between replicas. \\ \\ 

In order to concretely carry on the calculation we need
first to expand to the leading order the interaction term in the
partition function:
\bea
\exp\left\lbrace \beta J_a \sum_{\alpha=1}^x \prod_{i\in \partial a}\sigma_i^\alpha \right\rbrace 
\simeq \left( 1 + \beta J_a \sum_{\alpha=1}^x \prod_{i\in \partial a}\sigma_i^\alpha + \frac{1}{2}
\beta^2 J_a^2 \sum_{\alpha\beta=1}^x \prod_{i\in \partial a} \sigma_i^\alpha\sigma_i^\beta \right) 
\eea
By then indicating
\be
\int dm_{i\rightarrow a} Q_{i\to a}(m_{i\rightarrow a}) \prod_{\alpha=1}^x \eta_{i\rightarrow a}(\sigma_i^\alpha,m_{i\rightarrow a})~\mathcal{G}(\sigma_i^\alpha)= \left\langle \mathcal{G}(\sigma_i^\alpha) \right\rangle,
\ee
we have 
\bea
\beta\mathbb{F}_a^{ ^{\text{RSB}}}(x) &=& \log\left\lbrace  \left\langle e^{\beta J_a \sum_{\alpha=1}^x \prod_{i\in \partial a}\sigma_i^\alpha} \right\rangle \right\rbrace \simeq\nonumber \\
&\simeq& \log\left\lbrace  1 + \beta J_a \sum_{\alpha=1}^x \prod_{i\in \partial a}\langle \sigma_i^\alpha\rangle + \frac{1}{2}
\beta^2 J_a^2 \sum_{\alpha\beta=1}^x \prod_{i\in \partial a} \langle \sigma_i^\alpha\sigma_i^\beta\rangle \right\rbrace\simeq \nonumber \\
&\simeq& \beta J_a \sum_{\alpha=1}^x \prod_{i\in \partial a}\langle \sigma_i^\alpha\rangle + \frac{1}{2}
\beta^2 J_a^2 \sum_{\alpha\beta=1}^x \prod_{i\in \partial a} \langle \sigma_i^\alpha\sigma_i^\beta\rangle - \frac{1}{2}
\beta^2 J_a^2 \sum_{\alpha\beta=1}^x \prod_{i\in \partial a} \langle \sigma_i^\alpha\rangle \langle \sigma_i^\beta\rangle =\nonumber \\
&=& 
\gblue{ \beta J_a \sum_{\alpha=1}^x \prod_{i\in \partial a}\langle \sigma_i^\alpha\rangle +
\frac{1}{2} \beta^2 J_a^2 \sum_{\alpha=1}^x \prod_{i\in \partial a} \langle (\sigma_i^\alpha)^2\rangle +
\frac{1}{2} \beta^2 J_a^2 \sum_{\alpha\neq\beta}^x \prod_{i\in \partial a} \langle \sigma_i^\alpha\sigma_i^\beta\rangle
- \frac{1}{2} \beta^2 J_a^2 \sum_{\alpha\beta}^x \prod_{i\in \partial a} \langle \sigma_i^\alpha\rangle \langle \sigma_i^\beta\rangle} \nonumber \\
\eea

\gblue{ Finally, by making use of the definition of the moments $\langle
\sigma_i^\alpha \rangle$, $\langle (
  \sigma_i^\alpha)^2\rangle$ and $\langle
\sigma_i^\alpha\sigma_i^\beta\rangle$ given in
Eq.~(\ref{eq:1RSB-distribution-moments}) we can write
  \begin{align}
    \beta\mathbb{F}_a^{ ^{\text{RSB}}}(x) &= x \beta J_a  \prod_{i\in \partial a} h_{i\rightarrow a} + \frac{x}{2}\beta^2 J_a^2 \prod_{i\in \partial a}
    \left( \Delta^{(1)}_{i\rightarrow a} + \Delta^{(0)}_{i\rightarrow a} + h^2_{i\rightarrow a}\right) + \frac{x(x-1)}{2}\beta^2 J_a^2
    \prod_{i\in \partial a} \left(\Delta^{(0)}_{i\rightarrow a} + h^2_{i\rightarrow a}\right) - \frac{x^2}{2} J_a^2\prod_{i\in \partial a} h^2_{i\rightarrow a} \nonumber \\
  \end{align}

By then summing over all the $M$ interactions we can replace the local messages with their average values:
\begin{align}
  \sum_{a=1}^M \beta\mathbb{F}_a^{\text{RSB}}(x) = M x \left[ \beta \langle J \rangle \langle h \rangle^p + \frac{1}{2} \beta^2 \langle J^2\rangle
    + \frac{x-1}{2} \beta^2 \langle J^2 \rangle q_1^p - \frac{1}{2} \beta^2 \langle J^2 \rangle q_0^p \right].
\end{align}
If we assume the zero field case, i.e., $\langle J \rangle = \langle h \rangle = 0$, recalling then that $\langle J^2 \rangle = p!/(2N^{p-1})$
so that $M \langle J^2 \rangle \simeq N/2$ we finally obtain 

\be
\sum_{a=1}^M\beta\mathbb{F}_a^{ ^{\text{RSB}}}(x) = xN~\frac{\beta^2}{4} \left[ 1 - (1-x)q_1^p - x q_0^p\right]
\ee

}

\subsection{1RSB Entropy}
\label{sec:3D}
The 1RSB entropy is defined in eq. (\ref{eq:entropy-1rsb-Qm}) and we rewrite it here for the ease of the reader:
\be
\nonumber
\beta\mathbb{F}_i^{ ^{\text{RSB}}}(x) = \log \left\lbrace\int \prod_{ a\in\partial i}\prod_{k\in\partial a\backslash i}dm_{k\rightarrow a}~Q_{k\to a}(m_{k \rightarrow a})
~\hat{Z}_{a\to i}^x(\lbrace m_{k\rightarrow a}\rbrace_{k\in\partial a \setminus i})
e^{x \beta\mathbb{F}_i\left(\lbrace f(\lbrace m_{k\rightarrow a}\rbrace) \rbrace_{ a\in\partial i}\right)}
\right\rbrace
\ee
where the local free-energy $\mathbb{F}_i[\lbrace m_{k\rightarrow a}\rbrace_{k\in \partial a\setminus i}]$ is the RS one: 
\be
\exp\left(\beta\mathbb{F}_i\right) = \int_{-\infty}^\infty d\sigma_i~e^{-\lambda \sigma_i^2/2}\prod_{a\in\partial i}~
\left[\hat{Z}_{a\to i}^{-1}(\lbrace m_{k\rightarrow a}\rbrace_{k\in\partial a\setminus i})\int \prod_{k\in\partial a\setminus i}d\sigma_k~\eta^{\rm s}_{k\rightarrow a}(\sigma_k)~
e^{\beta J_a \sigma_i \prod_{k\in \partial a \setminus i}\sigma_k}\right].
\ee
By then assuming that $x$ is integer (the analytic continuation to
real values will be taken afterwards) it is straightforward to write:
\bea
&& \prod_{a\in\partial i}~\hat{Z}_{a\to i}^x(\lbrace m_{k\rightarrow a}\rbrace_{k\in\partial a\setminus i})~
e^{x\beta\mathbb{F}_i[\lbrace m_{k\rightarrow a}\rbrace_{k\in \partial a\setminus i}]}= \nonumber \\
&=& \int_{-\infty}^\infty \prod_{\alpha=1}^x d\sigma^\alpha_i~e^{-\lambda \sum_{\alpha=1}^x (\sigma_i^\alpha)^2/2}
\prod_{a\in\partial i}\left[\int \prod_{\alpha=1}^x \prod_{k\in\partial a\setminus i}d\sigma^\alpha_k
  ~\eta^{\rm s}_{k\rightarrow a}(\sigma^\alpha_k)~e^{\beta J_a \sum_{\alpha=1}^x \sigma_i^\alpha \prod_{k\in \partial a \setminus i}\sigma_k^\alpha}\right]
\eea
The implementation of the 1RSB ansatz comes at this stage very
natural, we just have to insert for each link $(ka)$  the ansatz in Eqns.~(\ref{eq:Q-1rsb-ansatz}),(\ref{eq:replicas-1rsb-ansatz}).
Thus, if we  take the average over  local fields
\emph{before} taking the one over  local variables---as for any replica calculation, but locally---it is convenient to define 
\be
\eta_{k\rightarrow a}(\boldsymbol{\sigma}_k) =
\int dm_{k\rightarrow a}~Q_{k\to a}(m_{k \rightarrow a})~\prod_{\alpha=1}^x \eta^{\rm s}_{k\rightarrow a}(\sigma_k^\alpha), 
\ee
which is precisely the quantity defined in Eq.~(\ref{eq:1RSB-ansatz}). The above steps allow us to rewrite
the 1RSB entropy simply as:
\bea
&& \beta\mathbb{F}_i^{\text{RSB}} = \log\left( \int_{-\infty}^\infty \prod_{\alpha=1}^x d\sigma^\alpha_i~e^{-\lambda \sum_{\alpha=1}^x (\sigma_i^\alpha)^2/2}
\prod_{a\in \partial i}
\left[\int_{-\infty}^{\infty} \prod_{k\in \partial a \setminus i} ~ \mathcal{D}\boldsymbol{\sigma}_k ~ \eta_{k\rightarrow a}(\boldsymbol{\sigma}_k) ~
  e^{\beta J_a \sum_{\alpha=1}^x \sigma_i^\alpha \prod_{k\in \partial a \setminus i}\sigma_k^\alpha} \right]\right) \nonumber \\
&=& \log\left( \int_{-\infty}^\infty \prod_{\alpha=1}^x d\sigma^\alpha_i~
e^{\left(\sum_{a\in\partial i} \hat{A}_{a\to i}\right)~\sum_{\alpha=1}^x \sigma_i^\alpha -\frac{1}{2}~\left(\lambda + \sum_{a\in\partial i} \hat{B}_{a\to i}^{(\text{d})}\right)~
\sum_{\alpha=1}^x (\sigma_i^\alpha)^2 + \frac{1}{2}~\left(\sum_{a\in\partial i} \hat{B}_{a\to i}^{(\text{nd})}\right)~ \sum_{\alpha\neq\beta}^x \sigma_i^\alpha\sigma_i^\beta} \right),\nonumber \\
\eea
where the quantities $\hat{A}_{a\to i}$, $\hat{B}_{a\to i}^{(\text{d})}$ and $\hat{B}_{a\to i}^{(\text{nd})}$
are those defined in Eq.~(\ref{eq:multiv-gauss-coeffs}).


\gblue{By comparing the definitions in
  Eq.~(\ref{eq:multiv-gauss-coeffs}) of the main text with those in
  Eq.~(\ref{eq:variable-function-equations}) and
  Eq.~(\ref{eq:variable-function-equations-2}) of
  Appendix~\ref{app:cavity-eq} one gets, at the leading order:
\bea
B^{(\text{d})} &=& \lambda+\sum_{a\in\partial i} \hat{B}_{a\to i}^{(\text{d})} \nonumber \\
B^{(\text{nd})} &=& \sum_{a\in\partial i} \hat{B}_{a\to i}^{(\text{nd})}, 
\eea
where the expressions of $B^{(\text{d})}$ and $B^{(\text{nd})}$ are
those given in Appendix~\ref{app:cavity-eq}.  According to the
definition of $\hat{A}_{a\rightarrow i}$ in
Eq.~(\ref{eq:multiv-gauss-coeffs}) it is then useful to define $A$ as follows:
\be
A = \sum_{a\in\partial i} \hat{A}_{a\rightarrow i} = \beta \sum_{a\in\partial i} J_a \prod_{j\in\partial a\setminus i} h_{j\rightarrow i}, 
\ee
from which one has
\be
A^2 = \beta^2 \langle J^2 \rangle \langle h^{2(p-1)} \rangle + \langle J \rangle^2 \langle h \rangle^{2(p-1)} =
\frac{q_0}{\left( 1- x q_0 - (1-x) q_1\right)^2},
\ee
where the last equality holds by virtue of Eq.~(\ref{eq:Asquared-2})
when $\langle J \rangle=\langle h \rangle=0$. }

We can, therefore, define a vector ${\bf A}$ 
 and a matrix $\mathcal{M}$ [already introduced in Eq.~(\ref{eq:matrixM})] as:
\bea
{A}_\alpha &=& A \ , \quad \forall \alpha  \nonumber \\
\mathcal{M}_{\alpha\beta} &=& \delta_{\alpha\beta} B^{(\text{d})}+ (1-\delta_{\alpha\beta}) (-B^{(\text{nd})}).
\nonumber\\
&& \alpha, \beta = 1,\ldots, x  \ . 
\label{eq:matrixM-reloaded}
\eea
The 1RSB entropy can be easily written as:
\bea
\beta\mathbb{F}_i^{\text{RSB}} &=& \log\left( \int \mathcal{D}\boldsymbol{\sigma}
\exp\left\lbrace \sum_{\alpha=1}^x A_\alpha\sigma_\alpha -
\frac{1}{2} \sum_{\alpha\beta} \sigma_\alpha \mathcal{M}_{\alpha\beta} \sigma_\beta \right\rbrace\right) \nonumber \\
&=& \sqrt{\frac{(2\pi)^x}{\det \mathcal{M}}}~
\exp\left( \frac{1}{2} A^2 \sum_{\alpha\beta} \mathcal{M}^{-1}_{\alpha\beta} \right)
\eea
where, making use of the definition of the inverse $\mathcal{M}^{-1}$
given in Eq.~(\ref{eq:inverse-M}) and with the help of a little
algebra we get:
\bea
\sum_{\alpha\beta} \mathcal{M}^{-1}_{\alpha\beta} &=& \frac{x}{B^{(\text{d})}+(1-x)B^{(\text{nd})}}
= \frac{x}{\frac{1}{1-q_1}- \frac{x(q_1-q_0)}{(1-q_1)(1-xq_0-(1-x)q_1)}} = x ~(1-xq_0-(1-x)q_1)
\eea
We thus have
\be
\frac{1}{2} A^2 \sum_{\alpha\beta} \mathcal{M}^{-1}_{\alpha\beta} = \frac{1}{2} \frac{\gblue{q_0}}{(1-xq_0-(1-x)q_1)^2}~\left(\sum_{\alpha\beta}\mathcal{M}^{-1}_{\alpha\beta}\right) =
\frac{xq_0}{1-xq_0-(1-x)q_1}
\ee

We also need to compute $\det \mathcal{M}$, that is the determinant of a symmetric
$x\times x$ matrix of the form:
\begin{equation*}
  \mathcal{M} = \left(\begin{array}{ccccc}
    a_1 & b & b & b & b  \\
    b & a_2 & b & b & b  \\
    b & b & a_3 & b & b  \\
    b & b & b & a_4 & b  \\
    b & b & b & b & a_5, 
  \end{array} \right) =
  \left(\begin{array}{ccccc}
    a_1-b & 0 & 0 & 0 & 0  \\
    0 & a_2-b & 0 & 0 & 0  \\
    0 & 0 & a_3-b & 0 & 0  \\
    0 & 0 & 0 & a_4-b & 0  \\
    0 & 0 & 0 & 0 & a_5-b, 
  \end{array} \right) +
  \left(\begin{array}{ccccc}
    b & b & b & b & b  \\
    b & b & b & b & b  \\
    b & b & b & b & b  \\
    b & b & b & b & b  \\
    b & b & b & b & b. 
  \end{array} \right)
\end{equation*}
The general formula for such a determinant is:
\be
\text{det}\left[ \text{Diag}(a_1-b,\ldots,a_x-b)+b~\cdot \mathbf{1}_x^T\otimes\mathbf{1}_x \right] = \prod_{i=1}^x(a_i-b)+b\sum_{i=1}^x\prod_{\substack{j=1 \\j\neq i}}^x (a_j-b).
\label{eq:determinant}
\ee
In the case of the rank-$x$ symmetric matrix $\mathcal{M}$ defined in
Eq.~(\ref{eq:matrixM-reloaded}), where in addition all the elements on
the diagonal are identical, from Eq.~(\ref{eq:determinant}) we have
that:
\be
\text{det}\left(\mathcal{M}\right) =\left( B^{(\text{d})} +  B^{(\text{nd})}\right)^x -  x~B^{(\text{nd})}~ \left( B^{(\text{d})} +  B^{(\text{nd})} \right)^{x-1}.
\ee
By exploiting then the definition of $B^{(\text{d})}$ and
$B^{(\text{nd})}$ in terms of $x$, $q_0$ and $q_1$ written in
Eq.~(\ref{eq:1RSB-matrix-elements}) one gets, after a very simple
algebra:
\be
\text{det}\left(\mathcal{M}\right) = \frac{1}{\left(1-q_1\right)^{x-1}}\cdot\frac{1}{\left(1-xq_0-(1-x)q_1\right)},
\ee
so that
\be
\frac{1}{\sqrt{\text{det}\left(\mathcal{M}\right)}} = \exp\left\lbrace \frac{1}{2} (x-1)\log\left( 1-q_1\right)+\frac{1}{2}\log\left( 1-xq_0-(1-x)q_1\right)\right\rbrace
\ee

At this point it is immediate to write
(neglecting constant terms) the 1RSB local free entropy term:
\be 
\sum_{i=1}^N\beta\mathbb{F}_i^{\text{RSB}} =N\left\{\frac{x}{2} \frac{q_0}{\left[ 1 - xq_0 - (1-x) q_1 \right]} + \frac{x-1}{2} \log\left( 1-q_1\right)+\frac{1}{2}\log\left[ 1-xq_0-(1-x)q_1\right]\right\}
\ee

\subsection{Vanishing of the normalization terms $\mathbb{F}_{ai}^{\text{RSB}}$}

Let us now show how the terms $\mathbb{F}_{ai}^{\text{RSB}}$ vanish in
the limit $N\rightarrow \infty$, provided that normalized distributions are
used everywhere. The first step is to plug the definition of
$\hat{Q}_{a\rightarrow i}(\hat{m}_{a \rightarrow i})$ of
Eq.~(\ref{eq:Qhat-def}) into the definition of
$\mathbb{F}_{ai}^{\text{RSB}}$ in Eq.~(\ref{eq:normalization-1rsb}),
thus obtaining:
\be
\exp\left[\beta\mathbb{F}_{ai}^{ ^{\text{RSB}}}(x)\right] = \int \left[\prod_{k\in\partial a \setminus i} d m_{k\rightarrow a} Q_{k\rightarrow a}(m_{k\rightarrow a})\right]~d{m}_{i\rightarrow a}~\hat{Z}_{a\to i}^x(\lbrace m_{k\rightarrow a}\rbrace)~Q_{i\to a}(m_{i\rightarrow a})~e^{x \beta\mathbb{F}_{ai}(m_{i \rightarrow a},\lbrace m_{k\rightarrow a}\rbrace )}.
\label{eq:Fai-1RSB-study}
\ee
From the RS equations Eqns.~(\ref{eq:cavity2}) and~(\ref{eq:free-energy-Zai}) we then have that:
\bea
e^{x \beta\mathbb{F}_{ai}(m_{i \rightarrow a},\lbrace m_{k\rightarrow a}\rbrace )} &=& Z_{(ai)}^x \nonumber \\
&=& \hat{Z}_{a\rightarrow i}^{-x}(\lbrace m_{k\rightarrow a}\rbrace) \left[\int_{-\infty}^\infty d\sigma_i~\eta_{i\rightarrow a}(\sigma_i)~e^{-\frac{\lambda \sigma_i^2}{2K}} \int_{-\infty}^\infty \prod_{k\in\partial a \setminus i} d\sigma_k~\eta_{k\rightarrow a}(\sigma_k)~e^{\beta J_a \sigma_i\prod_{k\in \partial a\setminus i}\sigma_k}\right]^x \nonumber \\
&=& \hat{Z}_{a\rightarrow i}^{-x}\left[ 1+\beta J_a \prod_{k\in \partial a} m_{k\rightarrow a}+\mathcal{O}(1/K)\right]^x.
\eea
Then, since we have $J_a \sim 1/N^{(p-1)/2}$ and $1/K \sim 1/N^{p-1}$, to the leading order in $N$ we can simply write
\be
x \beta\mathbb{F}_{ai}(m_{i \rightarrow a},\lbrace m_{k\rightarrow a}\rbrace ) =
- x \log\left[\hat{Z}_{a\rightarrow i}(\lbrace m_{k\rightarrow a}\rbrace)\right] + \mathcal{O}(1/N^{(p-1)/2}), 
\ee
so that inside the integral of Eq.~(\ref{eq:Fai-1RSB-study}) we have 
\be 
\hat{Z}_{a\rightarrow i}^{x}(\lbrace m_{k\rightarrow a}\rbrace) ~e^{x \beta\mathbb{F}_{ai}(m_{i \rightarrow a},\lbrace m_{k\rightarrow a}\rbrace )} \simeq 1. 
\ee

In conclusion, to the leading order in $N$, we can write
\be
\exp\left[\beta\mathbb{F}_{ai}^{ ^{\text{RSB}}}(x)\right] = \int \prod_{k\in\partial a} d m_{k\rightarrow a} ~ Q_{k\rightarrow a}(m_{k\rightarrow a}) = 1,
\ee
by the closure condition over the probability density $Q$.
 We have therefore shown that, to
the leading order in $N$, one has $\beta\mathbb{F}_{ai}^{ ^{\text{RSB}}}(x)=0$ for each edge $(ai)$.

\subsection{1RSB total free energy}
\label{sec:3E}

Putting together the results of
Sec.~\ref{sec:3C} and Sec.~\ref{sec:3D}, and remembering that the the
total free energy of the system is just the free energy of the $x$
coupled replicas that we computed, divided by $x$ (and extremized over
$x$), we obtain:
\bea
-x\beta\mathcal{F}(x) &=&
\sum_{a=1}^M \beta\mathbb{F}_a^{\text{RSB}} + \sum_{i=1}^N \beta\mathbb{F}_i^{\text{RSB}} -\sum_{(ai)\in E} \beta\mathbb{F}_{ai}^{\text{RSB}} \nonumber \\ 
&=& \frac{xN}{2} \Big\lbrace ~\frac{\beta^2}{2} \left[ 1 - (1-x)q_1^p - x q_0^p\right] + \frac{q_0}{\left[ 1 - xq_0 - (1-x) q_1 \right]} +
\frac{x-1}{x} \log\left( 1-q_1\right)+\frac{1}{x}\log\left[ 1-xq_0-(1-x)q_1\right] \Big\rbrace \nonumber \\
\eea

\bibliography{Giacomo}

\end{document}